\documentclass[fleqn,usenatbib,useAMS]{mnras}

\usepackage{newtxtext,newtxmath}

\usepackage[T1]{fontenc}
\usepackage{ae,aecompl}

\usepackage{graphicx}   
\usepackage{amsmath}    
\usepackage{amssymb}    
\usepackage{relsize}    

\renewcommand{\vec}[1]{\boldsymbol{#1}}
\newlength{\VSpaceBeforeTabBib}
\newlength{\VSpaceBeforeTabFoot}
\newcommand*\tablefootname{Notes}
\newcommand*\tablefootfont{\small}
\newcommand*\tablefootnamefont{\small\bfseries}
\newcommand\tablefoot[1]{\VSpaceBeforeTabBib=1ex%
  \par\vspace{\VSpaceBeforeTabFoot}
  \noindent
  \begin{minipage}{\linewidth}
    {\tablefootnamefont\tablefootname.}~%
    \tablefootfont
    \ignorespaces
    #1%
  \end{minipage}%
}
\newcommand*\tablefootmark[1]{%
  \unskip
  \hbox{\textsuperscript{\normalfont\itshape\ignorespaces#1}}%
  \,%
  \ignorespaces
}
\newcommand\tablefoottext[2]{%
  \hbox{\textsuperscript{\normalfont({\itshape\ignorespaces#1})}}%
  ~%
  \ignorespaces
  #2\ \ignorespaces%
}

\usepackage{xcolor}
\usepackage{xifthen}
\usepackage[normalem]{ulem}
\newcommand{\stkout}[1]{\ifmmode\text{\sout{\ensuremath{#1}}}\else\sout{#1}\fi}
\newcommand{\edited}[2]{\ifthenelse{\isempty{#1}}{\textcolor{red}{#2}}{\ifthenelse{\isempty{#2}}{\textcolor{gray}{\stkout{#1}}}{\textcolor{gray}{\stkout{#1}} \textcolor{red}{#2}}}}

\title[Magnetized Disk Formation]{Disk Formation in Magnetized Dense Cores with Turbulence and Ambipolar Diffusion}

\author[K. H. Lam et al.]{
Ka Ho Lam,$^{1}$\thanks{E-mail: kl4sf@virginia.edu}
Zhi-Yun Li,$^{1}$
Che-Yu Chen,$^{1}$
Kengo Tomida,$^{2,3}$
and Bo Zhao$^{4}$
\\
$^{1}$Department of Astronomy, University of Virginia, Charlottesville, VA 22904, USA\\
$^{2}$Department of Earth and Space Science, Osaka University, Toyonaka, Osaka 560-0043, Japan\\
$^{3}$Department of Astrophysical Sciences, Princeton University, Princeton, NJ 08544, USA\\
$^{4}$Max-Planck-Institut f\"ur extraterrestrische Physik (MPE), Garching D-85748, Germany
}

\date{Accepted XXX. Received YYY; in original form ZZZ}

\pubyear{2019}

\begin{document}
\label{firstpage}
\pagerange{\pageref{firstpage}--\pageref{lastpage}}
\maketitle

\begin{abstract}
Disks are essential to the formation of both stars and planets, but how they form in magnetized molecular cloud cores remains debated. This work focuses on how the disk formation is affected by turbulence and ambipolar diffusion (AD), both separately and in combination, with an emphasis on the protostellar mass accretion phase of star formation. We find that a relatively strong, sonic turbulence on the core scale strongly warps but does not completely disrupt the well-known magnetically-induced flattened pseudodisk that dominates the inner protostellar accretion flow in the laminar case, in agreement with previous work. The turbulence enables the formation of a relatively large disk at early times with or without ambipolar diffusion, but such a disk remains strongly magnetized and does not persist to the end of our simulation unless a relatively strong ambipolar diffusion is also present. The AD-enabled disks in laminar simulations tend to fragment gravitationally. The disk fragmentation is suppressed by initial turbulence. The ambipolar diffusion facilitates the disk formation and survival by reducing the field strength in the circumstellar region through magnetic flux redistribution and by making the field lines there less pinched azimuthally, especially at late times. We conclude that turbulence and ambipolar diffusion complement each other in promoting disk formation. The disks formed in our simulations inherit a rather strong magnetic field from its parental core, with a typical plasma-$\beta$ of order a few tens or smaller, which is 2-3 orders of magnitude lower than the values commonly adopted in MHD simulations of protoplanetary disks. To resolve this potential tension, longer-term simulations of disk formation and evolution with increasingly more realistic physics are needed.
\end{abstract}




\section{Introduction} \label{sec:intro}

Circumstellar disks play a central role in the formation of both Sun-like stars and planets. It is through such disks that the stars assemble most of their masses. The disk is also the birthplace for planets. Understanding the formation and evolution of disks has always been an integral part of the astronomical quest for our origins.

Despite significant progress, our knowledge of the origins of disks remains far from complete. A major impediment to a full understanding of how disks form and evolve is the magnetic field, which has been observed to thread molecular clouds in general \citep[e.g.,][]{2015A&A...576A.104P} and star-forming cloud cores in particular (for recent reviews, see \citealp{2019FrASS...6...15P} and \citealp{2019FrASS...6....3H}, and references therein). Since the magnetic field interacts closely with the movement of (partially) ionized gas, particularly the collapse and rotation of the magnetized star-forming core, it is expected to strongly affect the process of disk formation out of the dense core, although fully quantifying this effect remains a work in progress.

%
%
The potential for the magnetic field to strongly affect disk formation was demonstrated by early 2D (axisymmetric) numerical simulations of the collapse of magnetized rotating cores in the ideal MHD limit \citep[e.g.,][]{2000ApJ...528L..41T,2003ApJ...599..363A}, where the field removes essentially all of the angular momentum from the collapsing material through magnetic braking. \citet{2006ApJ...647..374G} pointed out that the efficient removal of angular momentum is directly tied to the well-known `magnetic flux problem' in star formation, namely, the stellar magnetic field would be many orders of magnitude stronger than the typically observed values if the magnetic flux threading the core was to be completely frozen into the matter and dragged all the way into the forming star, as would be the case in the strict ideal MHD limit. The concentration of magnetic flux at the center would formally lead to the formation of a split magnetic monopole in this limit, where both the rapid increase of the field strength toward the center and the long lever arm associated with the nearly radial field lines make the magnetic braking efficient and disk formation difficult. This difficulty was sometimes referred to as the `magnetic braking catastrophe' in the theoretical literature of disk formation. How the catastrophe is averted for disk formation is ultimately tied to how the magnetic flux problem is resolved in star formation.

The magnetic flux must be redistributed relative to the accreted matter in order to resolve the magnetic flux problem. The most studied means of flux redistribution is through non-ideal MHD effects, including Ohmic dissipation, ambipolar diffusion (AD) and the Hall effect. \citet{2006ApJ...647..382S} was the first to suggest that Ohmic dissipation may decouple the magnetic field from the circumstellar material enough to allow for disk formation. This suggestion was confirmed and extended numerically by \citet{2010ApJ...716.1541K} and \citet{2011PASJ...63..555M}, among others (see, e.g., \citealp{2014prpl.conf..173L} and \citealp{2016PASA...33...10T} for reviews of early work and references therein).

Non-ideal MHD effects are particularly well studied in recent years during the core collapse phase, up to (and slightly beyond) the formation of the second Larson's (stellar) core. In this early phase, there is now broad agreement that a relatively small disk (typically of several au or smaller in size) can form in the presence of Ohmic dissipation and ambipolar diffusion \citep[e.g.,][]{2012A&A...541A..35D,2015ApJ...801..117T,2015MNRAS.452..278T,2018A&A...615A...5V} and disk size can be increased or decreased by the Hall effect depending on whether the magnetic field is anti-aligned or aligned with the rotation axis \citep[e.g.,][]{2015ApJ...810L..26T,2018MNRAS.480.4434W}. The agreement is all the more remarkable in view of (1) the numerical challenges in covering the large range in spatial scale (from $>10,000$ to $\ll 1~\mathrm{au}$) and in treating non-ideal MHD effects and (2) the diverse techniques used in tackling the problem: semi-analytic \citep[e.g.,][]{2012A&A...541A..35D}, grid-based codes \citep[e.g.,][]{2015ApJ...801..117T,2018A&A...615A...5V}, and SPH codes \citep[e.g.,][]{2015MNRAS.452..278T,2015ApJ...810L..26T,2018MNRAS.480.4434W}.

%
%
How large (${\sim} 100~\mathrm{au}$ scale), persistent disks form and evolve during the later, main protostellar mass accretion phase of star formation is far less certain, as stressed by \citet{2016PASA...33...10T} and, more recently, \citet{2018MNRAS.473.2124G}. One potential difficulty in this phase is that, as more and more magnetized core material collapses onto the central protostellar system, more and more magnetic flux should be dragged by the collapsing material to the circumstellar region, making the magnetic field there increasingly stronger and magnetic braking increasingly more efficient, unless the magnetic flux can be effectively redistributed outward relative to the infalling matter. This redistribution of flux relative to matter lies at the heart of resolving the magnetic flux problem, which is much more severe at the end of the protostellar accretion phase than at the beginning, when the stellar mass is much larger and much more magnetic flux associated with the stellar mass needs to be redistributed outward. How it happens exactly is unclear, and is made more difficult by a technical challenge: simulations of the main protostellar accretion phase of star formation require a sink particle treatment (or something equivalent) to avoid the problem of prohibitively small time step shortly after the formation of the stellar seed.
%
%

Sink (or equivalent) treatment has been employed in magnetized protostellar disk formation simulations using both SPH \citep[e.g.,][]{2016MNRAS.457.1037W,2018MNRAS.477.4241L} and grid-based MHD codes. The latter include both ideal MHD simulations with turbulence \citep[e.g.,][]{2013MNRAS.432.3320S,2016ApJ...819...96G,2017ApJ...846....7K,2018MNRAS.473.2124G,2019MNRAS.486.3647K}, and non-ideal MHD simulations with Ohmic dissipation \citep[e.g.,][]{2011PASJ...63..555M,2014MNRAS.438.2278M,2017ApJ...835L..11T,2018A&A...620A.182K}, Ohmic dissipation and turbulence \citep[e.g.,][]{2017ApJ...839...69M}, ambipolar diffusion \citep[e.g.,][]{2016A&A...587A..32M,2016ApJ...830L...8H,2016MNRAS.460.2050Z,2018MNRAS.473.4868Z}, and all three non-ideal MHD effects but no turbulence \citep[e.g.,][]{2011ApJ...738..180L}. Ideally, one would want to include both turbulence and all three non-ideal MHD effects and follow the disk formation and evolution to the end of the protostellar mass accretion phase.

To achieve this goal, we have started a long-term program using the \textsmaller{Athena} MHD code \citep{2008ApJS..178..137S}. As a first step, we will focus on only one of the three non-ideal MHD effects, ambipolar diffusion, which has yet to be studied together with turbulence during the protostellar mass accretion phase using a sink treatment. We find both turbulence and ambipolar diffusion facilitate disk formation, but in a complementary way. The turbulence enables the formation of disks at early times and the ambipolar diffusion allows the turbulence-enabled early disks to persist to later times. In addition, the turbulence tends to make the AD-enabled disks less prone to gravitational fragmentation.

The rest of the paper is organized as follows. In \S~\ref{sec:setup}, we describe the setup of the numerical simulations. This is followed by a discussion of the results of the simulations that include only turbulence (\S~\ref{sec:turbulence}) and only ambipolar diffusion (\S~\ref{sec:ad}), respectively. We then discuss those disk formation simulations that include both turbulence and ambipolar diffusion in \S~\ref{sec:both}. Section \ref{sec:discussion} focuses on the gross properties of the formed disks, especially their degree of magnetization. The main results of the paper are summarized in \S~\ref{sec:conclusion}.

\section{Problem Setup}
\label{sec:setup}


\subsection{Governing Equations}

The non-ideal MHD equations including self-gravity that we solve numerically are:
\begin{equation}
  \frac{\partial \rho}{\partial t} + \nabla \cdot \left( \rho \vec{v} \right) = 0,
\end{equation}
\begin{equation}
  \label{eq:momentum}
  \rho \frac{\partial \vec{v}}{\partial t} + \rho \left( \vec{v} \cdot \nabla \right) \vec{v} = -\nabla P + \frac{1}{c} \vec{J} \times \vec{B} - \rho \nabla \Phi_g,
\end{equation}
\begin{equation}
  \frac{\partial \vec{B}}{\partial t} = \nabla \times \left( \vec{v} \times \vec{B} \right) - \frac{4 \pi}{c} \nabla \times \left( \eta_\mathrm{A} \vec{J}_\perp \right),
\end{equation}
\begin{equation}
  \nabla^2 \Phi_g = 4 \pi G \rho,
\end{equation}
where $\vec{J} = \left(c / 4 \pi \right) \nabla \times \vec{B}$ is the current density, and $\vec{J}_\perp = \left[ \left( \vec{J} \times \vec{B} \right) \times \vec{B} \right] / B^2$ is the current density perpendicular to the magnetic field. As a first step towards a comprehensive model, we make the simplifying assumptions that the gas is isothermal (with $P=\rho c_s^2$), that the ambipolar diffusivity $\eta_\mathrm{A}$ is given by
\begin{equation}
  \eta_\mathrm{A} = \frac{B^2}{4 \pi \gamma \rho \rho_i},
\end{equation}
where $\gamma = \left\langle \sigma v \right\rangle / \left( m + m_i \right)$ is the ion-neutral drag coefficient (charged grains are not accounted for explicitly), and that the ion density $\rho_i$ is approximated using $\rho_i = C \rho^{1/2}$, where $C$ is a constant \citep[assume equilibrium between cosmic-ray ionization and recombination, e.g.,][]{1992pavi.book.....S}; these approximations will be relaxed in future investigations.
Therefore, the ambipolar diffusivity can be rewritten into
\begin{equation}
  \eta_\mathrm{A} = Q_\mathrm{A} \frac{B^2}{4 \pi \rho^{3/2}}, \quad Q_\mathrm{A} = \frac{1}{\gamma C}.
  \label{eq:ADcoeff}
\end{equation}
The rest of the symbols have their usual meaning.

\subsection{Numerical Method}
We carry out a set of simulations in Cartesian coordinates using \textsmaller{Athena}, a grid-base code that solves 3D time-dependent non-ideal MHD equations including self-gravity \citep{2008ApJS..178..137S,2011ApJ...736..144B}. Roe solver is used for solving the MHD equations and self-gravity is solved using FFT with zero-padded boundary, which better isolates the core from its images, as if the computational domain is twice as big. Standard outflow boundary conditions are imposed in all three directions except for the velocity in the ghost zone, which is set to zero if it is pointing into the computational domain, to prevent material from entering the simulation box.
In order to speed up the simulations and follow the formation and evolution of disks for as long as possible, two treatments are employed.
\begin{enumerate}
  \item \textit{Sink particle} ---
  As mentioned in the introduction (\S~\ref{sec:intro}), we are interested in studying disk formation and evolution during the main protostellar accretion phase, where a sink particle treatment is necessary to avoid prohibitively small timesteps. Our implementation of sink particles is based on that by \citet{2013ApJS..204....8G}. We modified the original treatment slightly to better conserve mass and momentum.
  Specifically, the sink particle always lives in the center cell of a sink region of $3 \times 3 \times 3$ cells.
  Density and momentum thresholds are calculated for each cell in the sink region by averaging the closest neighbouring cells in the active zone.
  Any excess mass and momentum over the thresholds in the sink region are removed from the grid and put into the sink particle; the magnetic field in the sink region is left untouched.
  This implementation is intended to mimic the eventual decoupling of the material that is accreted onto the central protostar and the magnetic flux associated with it that must occur in order to resolve the well-known magnetic flux problem for the central star. While the treatment is a very rough approximation of the physics involved in the actual decoupling process, it does capture an essential aspect of the process, namely, the magnetic flux associated with the stellar material is not destroyed artificially\footnote{The magnetic flux in the densest disk-forming region is not destroyed by Ohmic dissipation or other non-ideal MHD effects \citep[as discussed in, e.g., the notes-added-in-proof of][]{2006ApJ...647..382S}. It is redistributed to lower density regions where the non-ideal MHD effects tend to be weaker.}; rather, it is preserved to machine accuracy since the magnetic field in the sink region is evolved in exactly the same manner as that in the active region using constrained transport (CT) in the \textsmaller{Athena} code.

%
%
  \item \textit{AD timestep floor} ---
  While running the non-ideal MHD simulations, we find the AD timestep often drops to very small values, causing the simulations to stall. From the definition
  \begin{equation}
    \label{eq:AD_timestep}
    \Delta t_\mathrm{AD} = \frac{{\Delta x}^2}{6 \eta_\mathrm{A}} = \frac{4 \pi \rho^{3/2} {\Delta x}^2}{6 Q_\mathrm{A} B^2},
  \end{equation}
  it is immediately clear that the AD timestep problem is most severe in cells with very low densities and moderately strong magnetic fields \citep{1995ApJ...442..726M}. We found that this is indeed the case. To speed up the simulations, we reduce the diffusivity in such cells locally by enforcing a lower limit (or floor) on the AD timestep. We monitor the affected cells to make sure that only a tiny amount of mass is affected by the treatment.
\end{enumerate}

\subsection{Model Setup and Parameters}
\label{sec:parameters}

We start our simulations with a $0.5~\mathrm{M_\odot}$ centrally-condensed spherical core with a radius of $2000~\mathrm{au}$ placed in a simulation box of $5000~\mathrm{au}$ on each side.
The density follows the pseudo-Bonner-Ebert sphere profile, which is described by
\begin{equation}
  \rho(r) = \frac{\rho_0}{1 + \left( r / r_c \right)^2},
\end{equation}
where $\rho_0$ is the central density and $r_c$ is the characteristic radius, which is chosen to be $1/3$ of the radius of the core so that the central density is $10 \times$ higher than the edge density.
The background density is set to one per cent of the edge density.
The isothermal sound speed $c_s$ is set to $0.2~\mathrm{km\,s^{-1}}$.
The core is assumed to have a solid-body rotation with an angular speed $\Omega \approx 6 \times 10^{-13}~\mathrm{s^{-1}}$ (corresponds to a rotational to gravitational energy ratio $\beta_\mathrm{rot} \approx 0.03$) with the rotational axis aligned with the $z$-axis.
This combination of parameters yields a large disk of ${\sim} 400~\mathrm{au}$ in the absence of a magnetic field that is easily resolvable in our simulations.
The magnetic field strength is characterized by the dimensionless mass-to-flux ratio $\lambda = 2 \pi \sqrt{G} \left( M_\mathrm{core}/\Phi \right)$, where $M_\mathrm{core}$ is the total mass of the core, $\Phi$ is the magnetic flux threading through the whole core, and $(2 \pi \sqrt{G})^{-1}$ is the critical value for the mass-to-flux ratio.
A uniform magnetic field along the rotation axis with strength corresponding to $\lambda \approx 2.6$ for the dense core as a whole is adopted in all of our simulations. We note that in our setup the mass-to-flux ratios along different (initially vertical) flux tubes are different, decreasing radially outward from a maximum value of ${\sim} 8.4$ on the axis. We also note that the above choice of dimensional numbers is not unique, since isothermal MHD simulations with self-gravity are scale-free. For example, one can choose a different length scale ($L$), which would lead to a corresponding change in the mass, density and magnetic field strength (in proportion to $L$, $L^{-2}$, and $L^{-1}$, respectively). The scale-free nature of the simulations is preserved in the presence of ambipolar diffusion with the adopted power-law dependence of the ion density on the neutral density ($\rho_\mathrm{i} = C\rho^{1/2}$).

In the models with turbulence, an initial turbulent velocity field is generated with an $k^{-2}$ power spectrum \citep{2011ApJ...729..120G}.
In order to ensure fair comparison between different models, any additional total angular momentum is removed from the turbulent velocity field before it is added on top of the solid-body rotation. The turbulence is allowed to decay freely after the initial injection. Since the dense core in our simulation collapses rather quickly, the majority of the turbulent motion is retained during the core collapse and disk formation.
The level of initial turbulence is characterized by the rms Mach number $\mathcal{M}$ such that an additional turbulent kinetic energy $E_\mathrm{turb} = M_\mathrm{core} \mathcal{M}^2 c_s^2 / 2$ is added. Three levels of turbulence are considered in this work, corresponding to $\mathcal{M} = 0.0$, $0.5$, and $1.0$, respectively. We do not consider supersonic turbulence because they are uncommon in low-mass star-forming cores \citep[e.g.,][]{2007ARA&A..45..339B}.

For ambipolar diffusion, we follow \citet{1992pavi.book.....S} and adopt ${\gamma = 3.5 \times 10^{13}~\mathrm{cm^3\,g^{-1}\,s^{-1}}}$ and ${C = 3 \times 10^{-16}~\mathrm{cm^{-3/2}\,g^{1/2}}}$ (corresponding to the standard cosmic ray ionization rate of $10^{-17}~\mathrm{s^{-1}}$), which are taken from \citet{1983ApJ...264..485D}. This combination yields a value of the coefficient for the ambipolar diffusivity $Q_{\mathrm{A}}$ defined in equation (\ref{eq:ADcoeff}) of $95.2~\mathrm{\,g^{1/2}\,cm^{-3/2}\,s}$,  which will be termed ``the standard value'' of the AD coefficient and denoted by $Q_{\mathrm{A},0}$ hereafter.  We will consider a range of $Q_\mathrm{A}$, including $0 \times$, $0.1 \times$, $0.3 \times$, $1 \times$, $3 \times$, and $10 \times$ the standard value. The model parameters are summarized in Table~\ref{tab:parameters}.

\begin{table}
  \caption{Model parameters and outcome}
  \label{tab:parameters}
  \begin{tabular}{lccl}
    \hline
    \hline
    Model Name & $\mathcal{M}$ & $Q_\mathrm{A}/Q_{\mathrm{A},0}$ & Comments \\
    \hline
    M0.0AD0.0  & 0.0 & 0.0  & DEMS\tablefootmark{a} \\
    M0.5AD0.0  & 0.5 & 0.0  & Transient Disk \& DEMS \\
    M1.0AD0.0  & 1.0 & 0.0  & Transient Disk \& DEMS \\
    M0.0AD0.1  & 0.0 & 0.1  & DEMS \\
    M0.0AD0.3  & 0.0 & 0.3  & DEMS \\
    M0.0AD1.0  & 0.0 & 1.0  & DEMS \& Persistent Disk\\
    M0.0AD3.0  & 0.0 & 3.0  & Persistent Disk \\
    M0.0AD10.0 & 0.0 & 10.0 & Persistent Disk \\
    M0.1AD0.1  & 0.1 & 0.1  & DEMS \\
    M0.1AD0.3  & 0.1 & 0.3  & DEMS \\
    M1.0AD0.1  & 1.0 & 0.1  & Transient Disk \& DEMS \\
    M1.0AD0.3  & 1.0 & 0.3  & Transient Disk \& DEMS \\
    M1.0AD1.0  & 1.0 & 1.0  & Transient Disk \& DEMS \\
    M1.0AD3.0  & 1.0 & 3.0  & Persistent Disk \\
    M1.0AD10.0 & 1.0 & 10.0 & Persistent Disk \\
    M0.0AD1.0US\tablefootmark{b} & 0.0 & 1.0 & DEMS \\
    M1.0AD1.0US\tablefootmark{b} & 1.0 & 1.0 & DEMS \\
    \hline
  \end{tabular}
  \tablefoot{
    \tablefoottext{a}{DEMS refers to the so-called `Decoupling-Enabled Magnetic Structure', a magnetically-dominated, low-density structure that is completely different from the dense rotationally supported disk \citep[see][and Fig.~\ref{fig:Turb_column_density} below]{2011ApJ...742...10Z}}
    \tablefoottext{b}{Simulation is initialized as a uniform sphere.}
  }
\end{table}

\subsection{Numerical Code and Zoom-in Simulations}

The \textsmaller{Athena} family of codes has been used for MHD simulations in a wide range of astrophysical systems, from clusters of galaxies \citep[e.g.,][]{2019MNRAS.483.2465M} to the atmospheres of planets \citep[e.g.,][]{2019ApJS..240...37L}. In star formation, it has been used to study the dynamics of magnetized molecular clouds and formation of dense cores and filaments \citep[e.g.,][]{2014ApJ...785...69C,2018ApJ...865...34C}. In this paper, we start a long-term program to extend such studies to core collapse and disk formation. This code choice is motivated by several factors. First, the code uses constrained transport (CT) to treat the magnetic field evolution, which ensures that the divergence-free condition $\nabla \cdot \vec{B} = 0$ is satisfied to the machine accuracy. This is of paramount importance for the magnetized disk formation problem because it prevents the generation of magnetic monopoles (and their associated change of magnetic field topology) even close to the forming protostar where a strong variation of the magnetic field is expected. Second, a well tested sink particle treatment including magnetic fields is already in place. Third, a treatment of ambipolar diffusion has already been implemented and has been applied successfully to the related problem of dense core formation \citep{2014ApJ...785...69C}.

A well-known difficulty with the treatment of ambipolar diffusion is that the time step required for numerical stability is proportional to the grid size ($\Delta x$) squared (see equation~\ref{eq:AD_timestep}). To alleviate this difficulty, we have decided to adopt a strategy of zoom-in simulations with uniform grids. Specifically, we adopt a base grid of $256^3$ for the pre-stellar phase of core evolution and restart each simulation right after the sink particle formation, keeping only the central $128^3$ cells in the original grid, and recasting them into $256^3$ cells as follows. Each of the kept original cells is split into eight octants of equal size. We keep the cell-centered hydrodynamics quantities untouched, which are later smoothed out as they evolve. Each of the three face-centered magnetic field components is linearly interpolated along its direction to each of the new faces, which ensures that the magnetic field on the new (finer) grid remains divergence-free. For our canonical choice of length scale, the resolution of the zoom-in simulation is ${\sim} 10~\mathrm{au}$. This relatively large cell size enables us to explore efficiently a wide range of simulation parameters, which is important for uncovering general trends, especially with respect to the strength of ambipolar diffusion. However, as discussed in \S~\ref{sec:parameters}, the physical scale of these simulations can in principle be reduced by choosing a smaller length scale.

We will first examine the trends in simulations where only turbulence or only ambipolar diffusion is included (\S~\ref{sec:turbulence} and \S~\ref{sec:ad}, respectively). This is followed by a discussion of those simulations where both effects are included (\S~\ref{sec:both}).


\section{Disk Formation in Ideal MHD: Turbulence}
\label{sec:turbulence}

To isolate the effects of turbulence from those of ambipolar diffusion, we will first concentrate on ideal MHD simulations with three different levels of turbulence characterized by $\mathcal{M} = 0.0$ (Model M0.0AD0.0; laminar), $0.5$ (M0.5AD0.0; subsonic), and $1.0$ (M1.0AD0.0; transonic).

\begin{figure*}
  \centering
  \includegraphics[width=\textwidth]{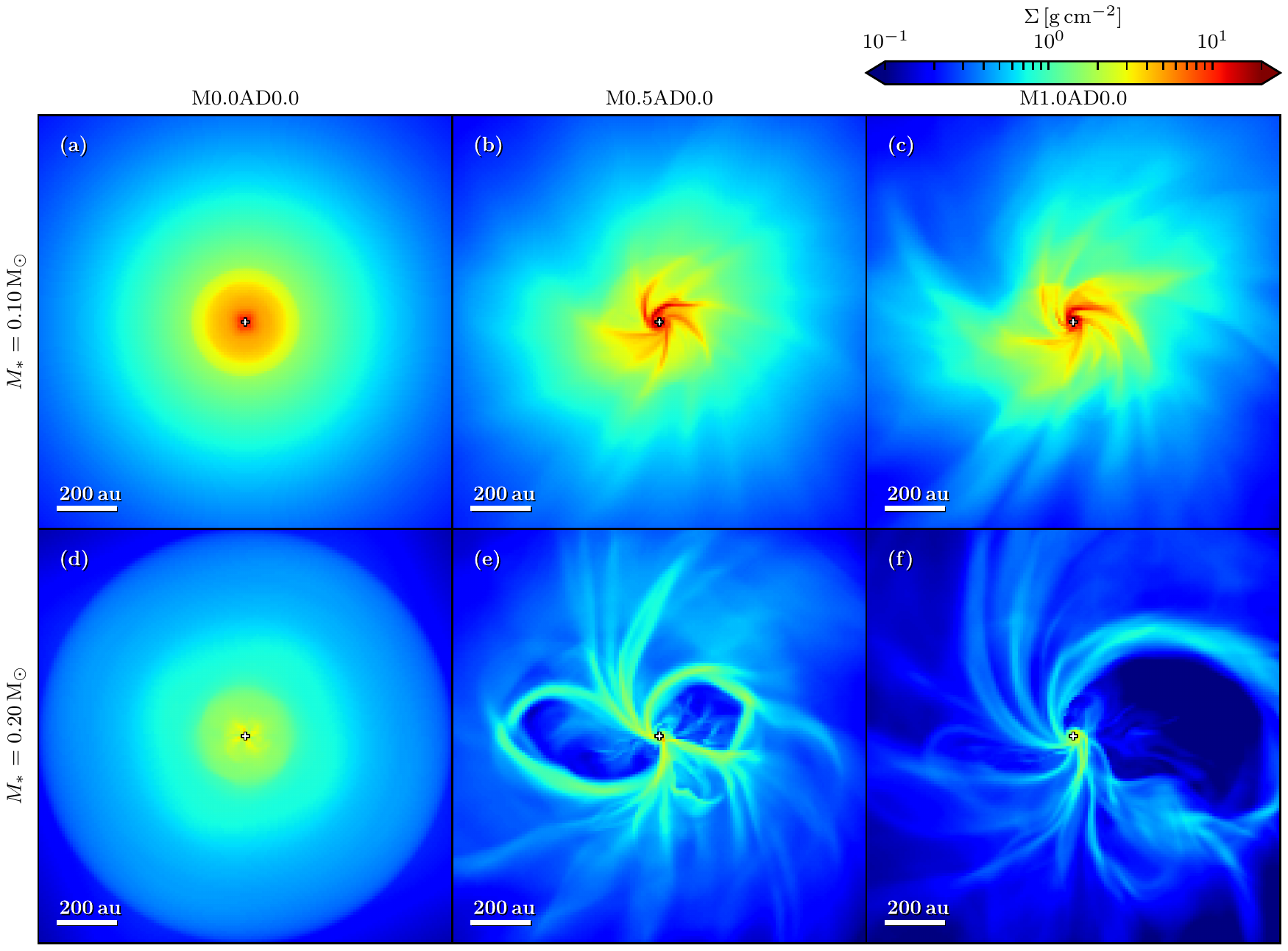}
  \caption{Column density along $z$-axis of the zoom-in simulations of the three ideal MHD models with different levels of turbulence (left to right, $\mathcal{M} = 0.0$, $0.5$ and $1.0$) when the sink particle has accreted $0.1~\mathrm{M_\odot}$ (upper row) and $0.2~\mathrm{M_\odot}$ (lower). The sink particle is marked by a cross. (See the supplementary material in the online journal for an animated version of the column density distribution for each model. Models M0.0AD0.0 and M1.0AD0.0 are also included in the animated version of Fig.~\ref{fig:AD_column_density} and \ref{fig:AT_column_density}, respectively.)}
  \label{fig:Turb_column_density}
\end{figure*}

Figure~\ref{fig:Turb_column_density} shows the column density along the $z$-axis (the rotation axis and the direction of the initial magnetic field) of the three models at two representative epochs when the sink particle has accreted $0.1~\mathrm{M_\odot}$ (upper row) and $0.2~\mathrm{M_\odot}$ (lower) of material, respectively. We choose to compare the models at the same sink (stellar) mass rather than the same absolute time (the time elapsed since the beginning of the simulation) or the relative time (the time elapsed since the formation of the sink particle) because, as mentioned earlier, the magnetic flux released from the stellar material plays a central role in the disk formation problem, and this flux is expected to be similar for the same stellar mass.

The simplest case of no turbulence (Model M0.0AD0.0) follows the well-known pattern that the gravitational collapse proceeds preferentially along the field lines, forming a thin, equatorial pseudodisk through which most of the material is accreted. Rotation of the infalling pseudodisk material winds up magnetic field lines, driving a bipolar outflow that removes angular momentum from the pseudodisk.
The magnetic flux brought in by accretion accumulates in the sink region and causes the formation of the so-called DEMS \citep[Decoupling-Enabled Magnetic Structure;][]{2011ApJ...742...10Z}, which persists until the end of the simulation. The combination of efficient angular momentum removal by outflow and the obstacle presented by the magnetically dominated DEMS makes it difficult to form a rotationally supported disk, which is absent from the laminar model.

In the presence of a subsonic (Model M0.5AD0.0) or transonic (M1.0AD0.0) turbulence, the basic picture remains broadly similar. In particular, both pseudodisk and DEMS still exist in the turbulent simulations. One difference is that the core collapse is slowed down somewhat by the additional kinetic energy associated with the turbulence. Another difference is that the pseudodisk is significantly warped, as discussed in more detail in \S~\ref{sec:turbulence_pseudodisk} below. Perhaps more importantly, the turbulence has induced the formation of prominent spiral structures close to the central protostar that are disk-like, at least at relatively early times (see panels b and c of Fig.~\ref{fig:Turb_column_density}). These disk-like structures are indicative of the beneficial effects of turbulence on disk formation. However, they largely disappear at later times as the circumstellar region becomes more dominated by DEMS (see panels e and f of Fig.~\ref{fig:Turb_column_density}). The effects of turbulence on disk formation will be discussed in more detail in \S~\ref{sec:turbulence_rotation} below.

We note that a large fraction, if not most, of the magnetic flux liberated from the central star is contained in the DEMS in these ideal MHD simulations, which is approximately the region where the plasma-$\beta$ is less than unity and the radial velocity is positive (i.e., expanding). For example, we have computed the magnetic flux threading such a region on the equatorial plane for the laminar ($\mathcal{M} = 0$) model when the stellar mass reaches $M_* = 0.25~\mathrm{M_\odot}$, and found a dimensionless ratio of the stellar mass to the DEMS magnetic flux of ${\sim} 4.8$. It is bracketed by the expected minimum value of ${\sim} 3.0$ (if the collapse is isotropic) and maximum of ${\sim} 5.4$ (if the stellar mass is accumulated along the field lines). Similarly, the ratio of the stellar mass to the DEMS magnetic flux is ${\sim} 4.9$ for the most turbulent model of $\mathcal{M} = 1$ when $M_* = 0.2~\mathrm{M_\odot}$ (see Fig.~1f), again consistent with the range between ${\sim} 3.1$ to ${\sim} 5.9$ expected for the stellar mass of $0.2~\mathrm{M_\odot}$.

\subsection{Warped Pseudodisk}
\label{sec:turbulence_pseudodisk}

In this subsection, we focus on the structure of the protostellar accretion flow on the several hundred to a few thousand au scale that is well resolved by our simulations. This region is important to study both theoretically and observationally. Theoretically, it is the bridge between the larger scale dense core and the smaller scale disk (if present). Observationally, it is starting to be probed by (sub)millimeter interferometers, especially ALMA. One may naively expect this region to be driven completely chaotic by turbulence. However, in the presence of a dynamically significant, large-scale magnetic field, the protostellar accretion flow remains spatially coherent to a large extent, as first demonstrated by \citet{2014ApJ...793..130L} for a non-self-gravitating accretion flow onto a star of fixed mass. Here we show that this basic result still holds when the self-gravity and a varying stellar mass are treated self-consistently.

\begin{figure*}
  \centering
  \includegraphics[width=\textwidth]{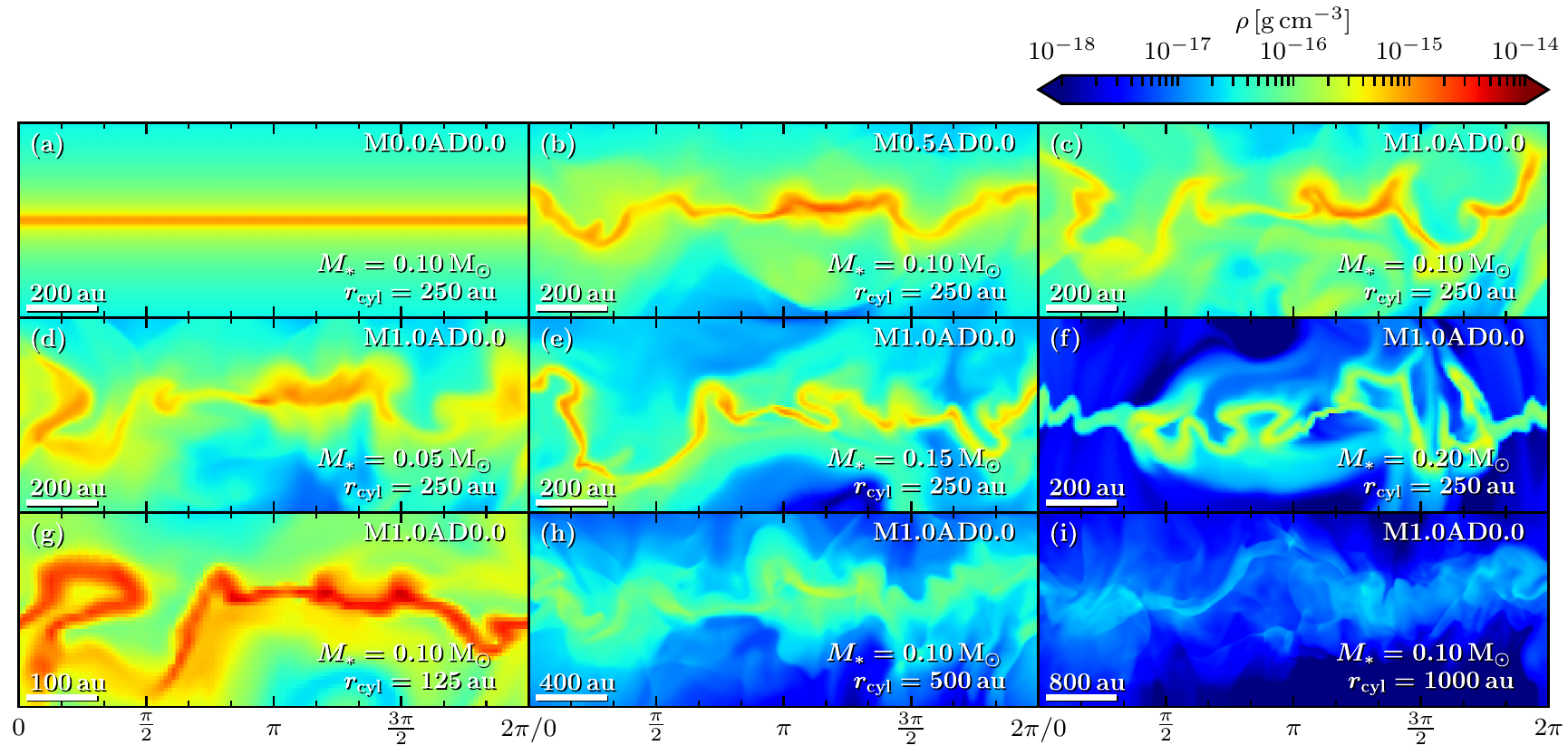}
  \caption{Turbulence-induced pseudodisk warping. Plotted in the top row are the density distributions on a cylinder of radius $r_\mathrm{cyl} = 250~\mathrm{au}$ at the epoch when $M_* = 0.1~\mathrm{M_\odot}$ for Model M0.0AD0.0 (panel a), M0.5AD0.0 (b), and M1.0AD0.0 (c) as a function of azimuthal angle $\phi$ (from 0 to $2\pi$) and height $z$, showing a more severe warping of the (dense) pseudodisk by a stronger turbulence. The middle row is for Model M1.0AD0.0 at the same radius as in panel (c) but at different epochs ($M_* = 0.05$ (d), $0.15$ (e), and $0.2~\mathrm{M_\odot}$ (f)), showing the time evolution of the warped pseudodisk. The bottom row is also for Model M1.0AD0.0 at the same epoch as in panel (c) but at different cylindrical radii ($125$ (g), $500$ (h), and $1000~\mathrm{au}$ (i)).}
  \label{fig:Turb_density_on_cylinder}
\end{figure*}

\begin{figure}
  \centering
  \includegraphics[width=\columnwidth]{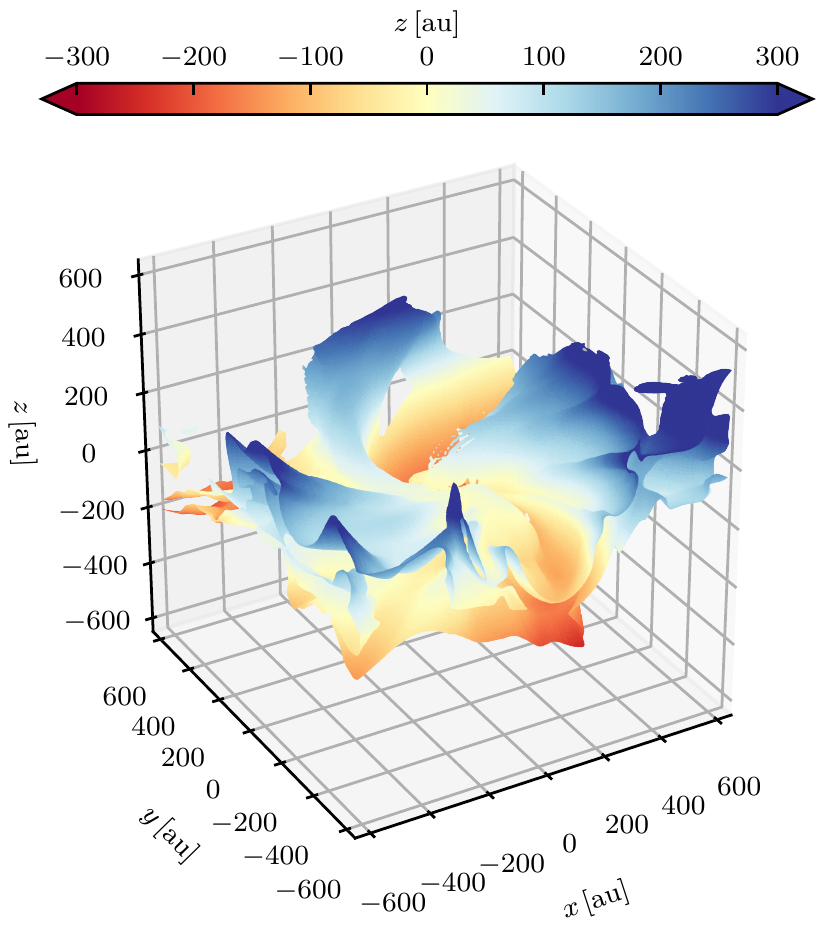}
  \caption{3D view of the turbulence-warped pseudodisk. Plotted is the isosurface of the normalized density $\tilde{\rho} = 1$ at an epoch when the stellar mass $M_* = 0.15~\mathrm{M_\odot}$ for the ideal MHD sonic turbulence model M1.0AD0.0. The surface is colored by its height above or below the $x$-$y$ plane passing through the sink particle (i.e., its $z$ value).}
  \label{fig:Turb_3d_pseudodisk}
\end{figure}

We demonstrate the coherence of the density structure in two ways, through density distributions on cylindrical surfaces around the $z$-axis passing through the sink particle (Fig.~\ref{fig:Turb_density_on_cylinder}) and three-dimensional visualization (Fig.~\ref{fig:Turb_3d_pseudodisk}). Panel (a) of Fig.~\ref{fig:Turb_density_on_cylinder} shows that, at a representative epoch when $M_* = 0.1~\mathrm{M_\odot}$, the density distribution of the non-turbulent model (M0.0AD0.0) on a representative cylinder of radius $r_\mathrm{cyl} = 250~\mathrm{au}$ is concentrated near the equator. This is of course the well-known pseudodisk. As the level of turbulence increases, the pseudodisk becomes increasingly more warped (compare panels a-c) but remains spatially connected. This coherence persists to later times, even in the case of strongest turbulence (M1.0AD0.0; see panels d-f, where the density distributions are plotted for the epochs when $M_* = 0.05$, $0.15$ and $0.2~\mathrm{M_\odot}$, respectively). The pseudodisk warping is not limited to the particular radius of $r_\mathrm{cyl} = 250~\mathrm{au}$, as illustrated in panels (g)-(i), where we plot the density distributions for Model M1.0AD0.0 at the same time as shown in panel (c) but on cylinders of three other radii ($r_\mathrm{cyl} = 125$, $500$ and $1000~\mathrm{au}$).

%

The spatial coherence of the turbulence-warped pseudodisk can be visualized more clearly in Fig.~\ref{fig:Turb_3d_pseudodisk}, which shows the isosurface of the normalized density
\begin{equation}
  \tilde{\rho} = \frac{\rho r_\mathrm{cyl}}{\Sigma} = 1,
\end{equation}
where $r_\mathrm{cyl}$ is the cylindrical radius and $\Sigma$ is the column density along the direction of the initial magnetic field ($z$-axis). Physically, $\tilde{\rho}$ is a dimensionless quantity that is the inverse of the characteristic thickness of the local density structure ($\Sigma/\rho$) relative to the local cylindrical radius. It is a measure of the (angular) `thinness' of the density structure. We find it easier to highlight regions with high mass concentration at different radii simultaneously (i.e., the pseudodisk) using this dimensionless quantity than the density itself, because the latter varies much more strongly with radius.

The 3D view of the pseudodisk drives home the important conceptual point that, in the presence of a dynamically significant large-scale magnetic field and a subsonic or transonic turbulence, the inner protostellar accretion flow has a unique texture that is neither completely chaotic (as expected, e.g., for a region of isotropic turbulence) nor simply organized (as the equatorial pseudodisk in the laminar Model M0.0AD0.0). Instead, it has a structure intermediate between these two extremes that is shaped by the interaction between the large-scale magnetic field and gravity, which tends to produce a flattened structure (i.e., a pseudodisk) because of the anisotropy in the magnetic support against the gravity, on the one hand, and by the turbulent motions initially present inside the core, which tend to perturb the flattened structure by deflecting the gravity-induced collapsing motions towards the pseudodisk and by distorting the magnetic field lines. The perturbed pseudodisk is further modified by rotation, especially at small radii where the rotational speed is typically the highest. Indeed, the spiral structures that are prominent in the column density maps of the turbulent models in Fig.~\ref{fig:Turb_column_density} are those more tilted parts of the warped pseudodisk that are viewed more edge-on (i.e., with a longer path length in the pseudodisk along the line of sight), as already discussed in \citet{2014ApJ...793..130L} in the absence of self-gravity. The inclusion of self-gravity in our simulations strengthens the general picture that the warped pseudodisk serves as a dense `backbone' for the inner protostellar accretion flow onto low-mass stars formed in turbulent, magnetized dense cores. It would be interesting to test this picture through high-resolution observations, especially using ALMA.

%
%


\subsection{Enhanced Rotation by Turbulence}
\label{sec:turbulence_rotation}

\begin{figure*}
  \centering
  \includegraphics[width=\textwidth]{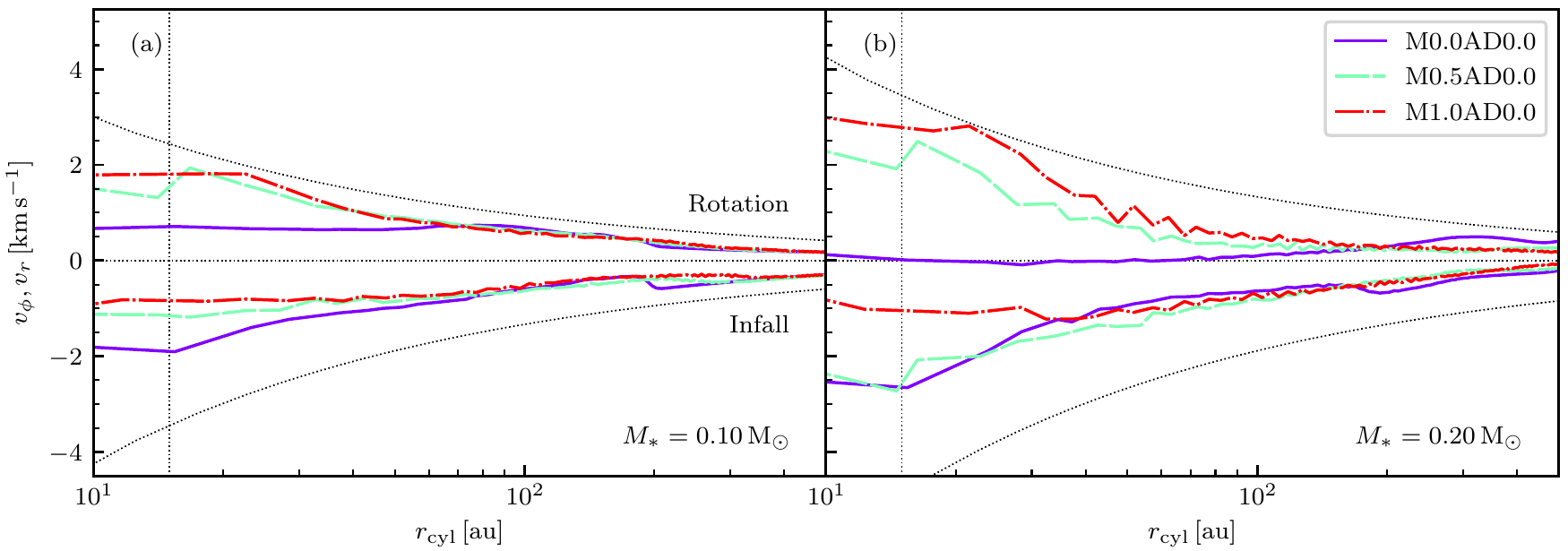}
  \caption{Distributions of the mass-weighted rotational (upper curves) and infall speeds (lower) in a wedge within $45^\circ$ of the equatorial plane compared to the Keplerian (upper black dotted line) and free-fall (lower) profile at two representative epochs with stellar mass of $M_* = 0.1$ (panel a) and $0.2~\mathrm{M_\odot}$ (b) for three ideal MHD models with turbulent Mach number of $\mathcal{M} = 0.0$ (violet solid line), $0.5$ (green dashed) and $1.0$ (red dash-dotted). The vertical dotted line in each panel denotes roughly the radius of the sink region.}
  \label{fig:Turb_vel}
\end{figure*}

In Fig.~\ref{fig:Turb_column_density}, we have already seen hints of the beneficial effects of turbulence on disk formation from the morphology of the circumstellar material. These effects are quantified in Fig.~\ref{fig:Turb_vel}, which plots the mass-weighted distributions of the infall and rotational speeds as a function of radius at the same two representative stellar mass $M_* = 0.1$ and $0.2~\mathrm{M_\odot}$ as in Fig.~\ref{fig:Turb_column_density}. At the relatively early epoch when $M_* = 0.1~\mathrm{M_\odot}$, the bulk of the circumstellar material on the 100-au scale rotates well below the Keplerian speed in the laminar case (Model M0.0AD0.0), with a rotational speed significantly below the infall speed, which is indicative of a rapidly collapsing inner protostellar envelope rather than a rotationally supported structure. This is in contrast with the turbulent cases where the rotational speed is significantly closer to the Keplerian value and the infall speed closer to zero. This is especially true for the sonic turbulence model (M1.0AD0.0) where the rotational speed approaches the Keplerian speed outside the sink region ($r_\mathrm{cyl} \gtrsim 15~\mathrm{au}$), and the infall speed is ${\sim} 2-3$~times lower than that (and well below the free-fall value). Whether this slowly-collapsing (compared to free fall), rotationally-dominated, flattened circumstellar structure is called a `disk' or not depends on how disks are defined \citep[e.g.,][]{2018A&A...615A...5V}. A quantitative definition of disks will be described below in \S~\ref{sec:2disk}.

The sonic turbulence model (M1.0AD0.0) is strongly affected by the magnetically-dominated, low-density DEMS at later times. For example, a well-developed DEMS is clearly visible in the surface density plot of Fig.~\ref{fig:Turb_column_density} when $M_* = 0.2~\mathrm{M_\odot}$ (see panel f). Nevertheless, the infall speed remains well below the free-fall value even at this late epoch, and the (mass-weighted) rotational speed still approaches the Keplerian value right outside the sink region, as illustrated by the red curves in panel (b) of Fig.~\ref{fig:Turb_vel}. At this epoch, the appearance of the circumstellar region of the weaker turbulence case (M0.5AD0.0) is also dominated by DEMS, although its rotational speed is somewhat lower on average and its infall motion somewhat faster compared to Model M1.0AD0.0 (compare green and red curves in panel b). This trend continues to the laminar case (violet curves), especially for the rotational speed, which is close to zero on the 100-au scale. This comparison re-enforces the notion that turbulence facilitates disk formation, making it possible for the circumstellar material to rotate closer to the Keplerian speed.

There are several reasons suggested for why turbulence tends to promote disk formation. They include turbulence-induced magnetic diffusion \citep[e.g.,][]{2012ApJ...747...21S}, tangling of magnetic field lines \citep[e.g.,][]{2013MNRAS.432.3320S}, field-rotation misalignment \citep[e.g.,][]{2013A&A...554A..17J,2018MNRAS.473.2124G}, and pseudodisk warping \citep[e.g.,][]{2014ApJ...793..130L}. To these we add two more possibilities: earlier formation of DEMS and self-sorting of infalling protostellar envelope materials of different specific angular momenta.

\begin{figure*}
  \centering
  \includegraphics[width=\textwidth]{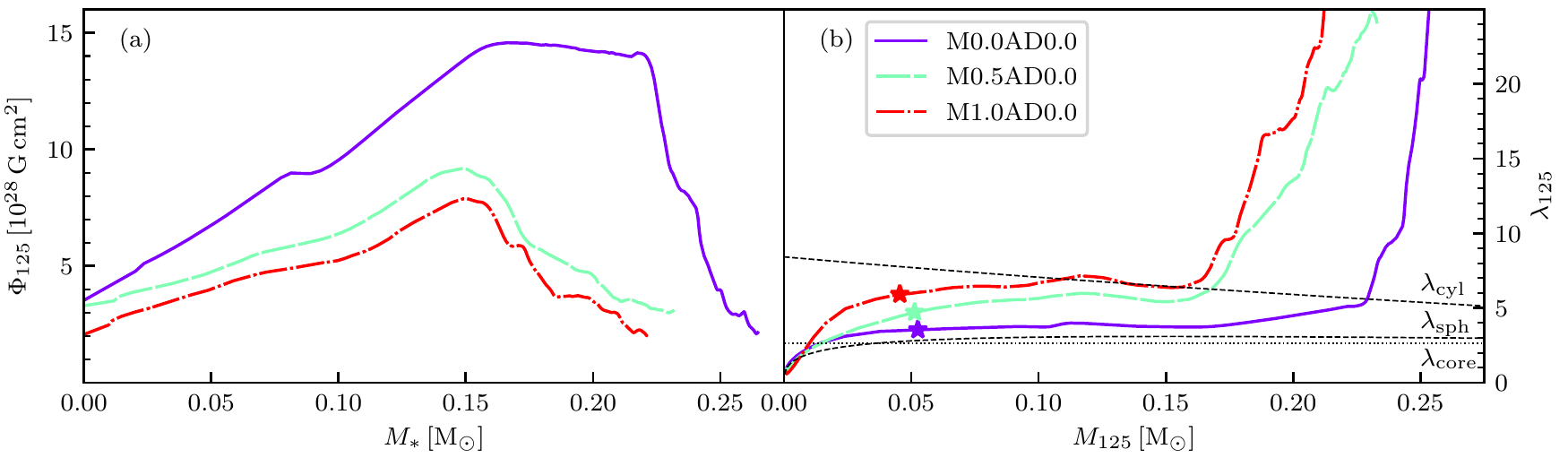}
  \caption{Reduction of the magnetic flux close to the accreting protostar by turbulence. Plotted are (a) the magnetic flux passing through a circle of $125~\mathrm{au}$ in radius on the equatorial plane as a function of stellar mass and (b) the dimensionless ratio of the mass enclosed by a sphere of $125~\mathrm{au}$ in radius (including the stellar mass) and the magnetic flux as a function of the enclosed mass for ideal MHD models of turbulence Mach number $\mathcal{M} = 0.0$ (violet solid line), $0.5$ (green dashed), and 1.0 (red dash-dotted). The dotted line in panel (b) denotes the initial value for the core as a whole $\lambda_\mathrm{core}$ and the upper and lower dashed lines denote, respectively, the mass-to-flux ratio expected under flux freezing in the limit that the stellar mass is accumulated along the (initially vertical) magnetic field lines ($\lambda_\mathrm{cyl}$) and that the core collapse into the star is strictly spherical or isotropic ($\lambda_\mathrm{sph}$). The star symbol denotes the time of sink particle formation for each model.}
  \label{fig:Turb_Bflux}
\end{figure*}

Strongly magnetized, low-density, expanding DEMS are formed in all three ideal MHD cases (see e.g. Fig.~\ref{fig:Turb_column_density}) especially at the later epoch when $M_* = 0.2~\mathrm{M_\odot}$. It turns out that such DEMS form at a smaller stellar mass for the turbulent case compared to the laminar case, as can be seen most clearly from the animation of the column density map in the supplementary material of the online version of the article. The reason for the earlier (in terms of stellar mass) formation is that DEMS are produced by a competition between the ram pressure of the infalling material and the magnetic forces exerted by the magnetic flux decoupled from the accreted mass.
As the star accretes mass, the magnetic flux associated with the accreted mass is left in its surroundings, causing an accumulation of magnetic flux. Although the ram pressure increases initially due to the increasing central mass and infall speed, it drops with the density at later times. When the ram pressure becomes dominated by the magnetic forces, DEMS form. This process is facilitated by turbulence, which produces inhomogeneity in the density distribution in the circumstellar gas, including low-density channels between dense spirals where the trapped magnetic flux can leak out more easily and, therefore, at an earlier epoch. The flux leakage lowers the magnetic flux near the accreting protostar, as illustrated in the panel (a) of Fig.~\ref{fig:Turb_Bflux}, which shows a much lower magnetic flux passing through an equatorial circle of $125~\mathrm{au}$ in radius centered on the protostar for each of the two turbulence models compared to the laminar model. The lower magnetic flux in turn promotes disk formation.

There is the possibility that the decrease of magnetic flux in the central region with increasing turbulence is due to enhanced magnetic diffusion from turbulence-induced reconnection \citep[as reviewed by][]{2015RSPTA.37340144L}. Indeed, in their ideal MHD simulations of strongly magnetized, turbulent, cluster-forming clouds, \citep{2015MNRAS.452.2500L} found that the mass-to-flux ratios of the most massive dense clumps are often larger than that of the cloud as a whole, indicating a breakdown of the flux-freezing condition, possibly due to turbulent reconnection. In panel (b) of Fig.~\ref{fig:Turb_Bflux}, we plot the dimensionless ratio $\lambda_{125}$ of the mass $M_{125}$ enclosed within a sphere of $125~\mathrm{au}$ in radius (including the stellar mass) to the magnetic flux passing through an equatorial circle of $125~\mathrm{au}$ (shown in panel a) as a function of the enclosed mass, and compare it to the initial mass-to-flux ratio of the core as a whole ($\lambda_\mathrm{core}$, the dotted line in the panel), as well as the mass-to-flux ratios expected under flux freezing in two limits: (1) the stellar mass is accumulated along the (initially vertical) magnetic field lines ($\lambda_\mathrm{cyl}$, upper dashed line), and (2) the core collapse into the star is strictly spherical or isotropic ($\lambda_\mathrm{sph}$, lower dashed line). It is clear that although $\lambda_{125}$ is significantly larger than $\lambda_\mathrm{core}$ and $\lambda_\mathrm{sph}$ over most of the time for the turbulent models, it remains near or below $\lambda_\mathrm{cyl}$, except towards the end of the simulation, when $\lambda_{125}$ increases rapidly above $\lambda_\mathrm{cyl}$. Therefore, the relatively high values of $\lambda_{125}$ before its rapid rise could in principle come from mass accumulation along field lines rather than turbulent reconnection, although some contribution from the latter cannot be excluded. The rapid increase in $\lambda_{125}$ towards the end is due to rapid expansion of DEMS, which happens even for the laminar case without any turbulence.

Another way that turbulence can help with disk formation is that, unlike the laminar case with a flat equatorial pseudodisk, the bulk of rotating, protostellar material can fall close to the central protostar on different planes (see Fig.~\ref{fig:Turb_3d_pseudodisk}).
In the presence of a strong magnetic field, material on the same magnetic field line would tend to flow along the field line and form the dense (warped) pseudodisk, as illustrated in Figure~\ref{fig:Turb_density_on_cylinder} and discussed in \S~\ref{sec:turbulence_pseudodisk}.  The turbulence-induced variation of angular momentum of the material initially located on the same field line would be largely erased once the bulk of this material has collapsed along the field line onto the pseudodisk. Nevertheless, there can still be variation of angular momentum between the materials collected onto the pseudodisk along different field lines. Because of strong warping, different parts of the pseudodisk with different specific angular momenta can fall towards the central protostar on different planes, which reduces their chance of collision. This makes it easier for the high specific angular momentum material to retain its angular momentum and form a disk, without being impeded by the low specific angular momentum material, which can fall into the sink region without colliding with the high specific angular material.

\section{Disk Formation in Non-ideal MHD: Ambipolar Diffusion}
\label{sec:ad}

In this section, we seek to isolate the effects of ambipolar diffusion by setting the turbulence to zero and considering a wide range of ambipolar diffusivity, with the coefficient $Q_\mathrm{A} = 0 \times$ (Model M0.0AD0.0), $0.1 \times$ (M0.0AD0.1), $0.3 \times$ (M0.0AD0.3), $1 \times$ (M0.0AD1.0), $3 \times$ (M0.0AD3.0), and $10 \times$ (M0.0AD10.0) the standard value. The choice of the largest AD coefficient is motivated by \citet{2016MNRAS.460.2050Z,2018MNRAS.473.4868Z}, who showed that the level of ambipolar diffusion can be enhanced by 1-2 orders of magnitude when small grains are depleted. We will first survey the broad trends (\S~\ref{sec:AD_trends}) before diving into detailed discussions of how ambipolar diffusion affects the protostellar accretion flow relative to the ideal MHD case (\S~\ref{sec:AD_ideal}) and the reasons behind the trends identified (\S~\ref{sec:AD_diskform}).

\subsection{Overview of Results}
\label{sec:AD_trends}

\begin{figure*}
  \centering
  \includegraphics[width=\textwidth]{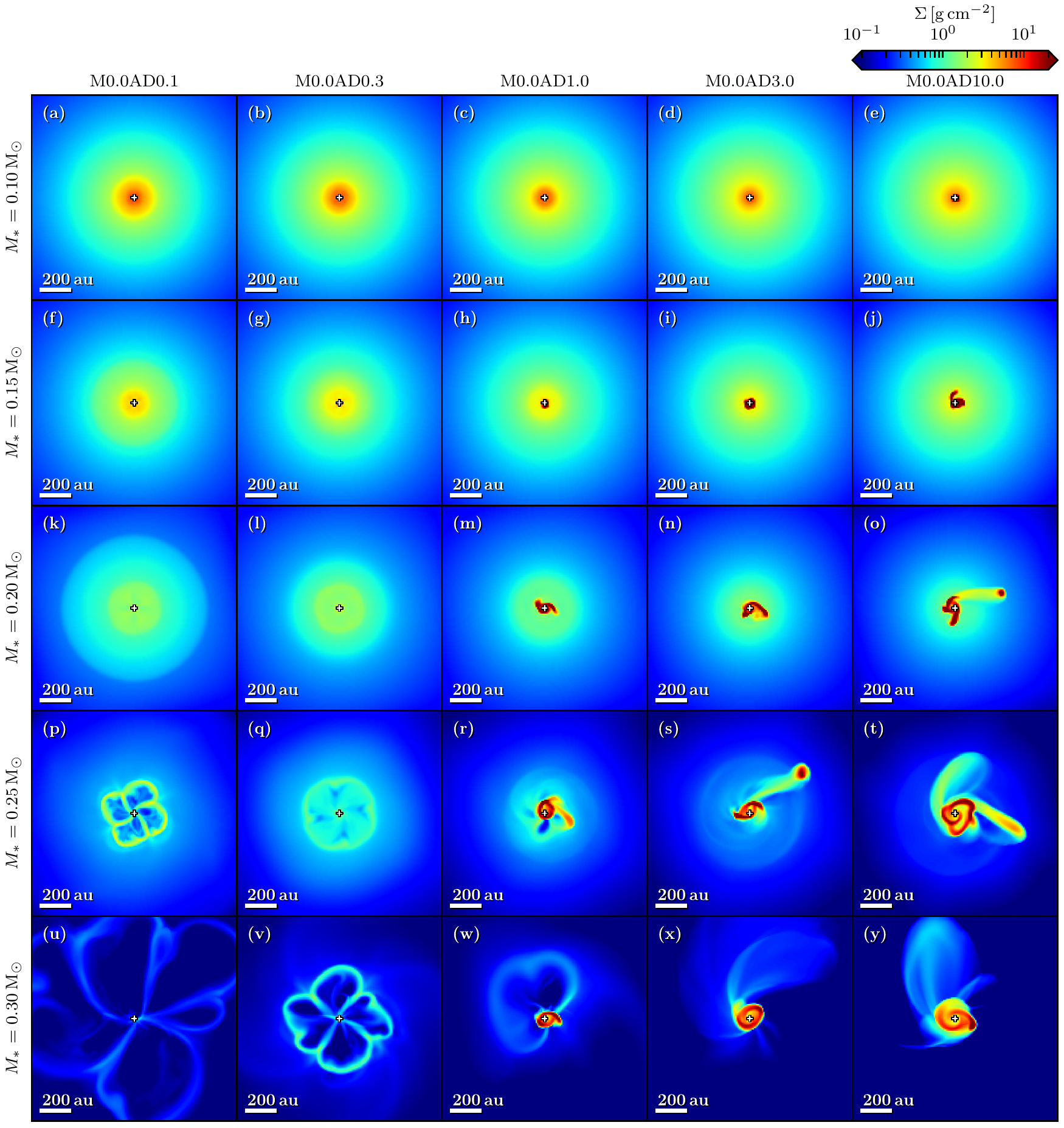}
  \caption{Column density along $z$-axis of the zoom-in simulations of all non-turbulent AD models with $Q_\mathrm{A} = 0.1 \times$, $0.3 \times$, $1.0 \times$, $3.0 \times$ and $10.0 \times$ (left to right) the standard value when the sink particle has accreted $0.1$, $0.15$, $0.2$, $0.25$ and $0.3~\mathrm{M_\odot}$ (top to bottom). The sink particle is marked by a cross. (See the supplementary material in the online journal for an animated version of the column density distribution for each model.)}
  \label{fig:AD_column_density}
\end{figure*}
\begin{figure*}
  \centering
  \includegraphics[width=\textwidth]{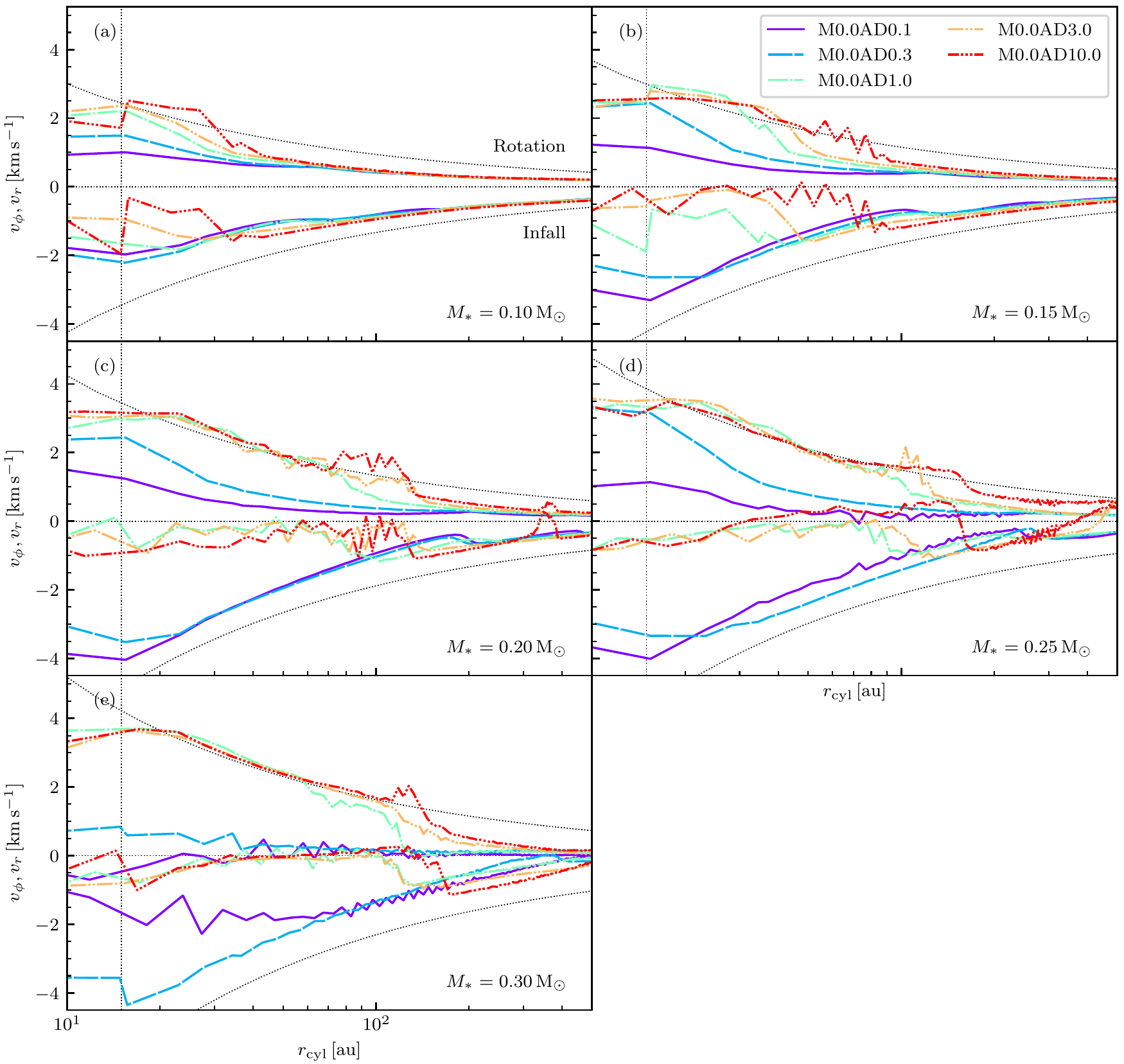}
  \caption{Distributions of the mass-weighted rotational (upper curves) and infall speeds (lower) on the equatorial plane compared to the Keplerian (upper black dotted) and free-fall (lower) profile at five representative epochs with stellar mass of $M_* = 0.1$ (panel a), $0.15$ (b), $0.2$ (c), $0.25$ (d) and $0.3~\mathrm{M_\odot}$ (e) for all non-turbulent AD models with $Q_\mathrm{A} = 0.1 \times$ (violet solid line), $0.3 \times$ (blue dashed), $1.0 \times$ (green dash-dotted), $3.0 \times$ (yellow dash-double-dotted) and $10.0 \times$ (red dash-triple-dotted) the standard value. The vertical dotted line in each panel denotes roughly the radius of the sink region.}
  \label{fig:AD_vel}
\end{figure*}

To get a first impression on how ambipolar diffusion affects the dynamics of core collapse and disk formation, we plot in Fig.~\ref{fig:AD_column_density} and \ref{fig:AD_vel}, respectively, the column density maps and the mass-weighted infall and rotational speeds as a function of radius for the material on the the equatorial plane for all non-turbulent AD models (rather than within a wedge of 45$^\circ$ of the equatorial plane as in Fig.~\ref{fig:Turb_vel} since the pseudodisk here is not warped by turbulence) at five epochs when the stellar mass $M_* = 0.1$, $0.15$, $0.2$, $0.25$, and $0.3~\mathrm{M_\odot}$. From the left-most column of the column density maps, it is clear that the least magnetic diffusive model, M0.0AD0.1, does not show any evidence of a well-formed disk, especially at later epochs, when the appearance of the circumstellar region is dominated by low-density, expanding regions (i.e., DEMS). The lack of a rotationally supported structure is corroborated by the velocity profiles displayed as violet solid curves in the Fig.~\ref{fig:AD_vel}, which show that the rotation is significantly sub-Keplerian on the 100-au scale at the earliest epoch (when $M_* = 0.1~\mathrm{M_\odot}$) and becomes worse at later times, and that the rotational speed is increasingly dominated by the infall speed over time.

As the AD coefficient $Q_\mathrm{A}$ increases from 0.1 to 0.3 times the standard value, the (mass-weighted) rotational speed stays closer to the Keplerian value (compare blue dashed and violet solid curves in Fig.~\ref{fig:AD_vel}) until the last epoch (when $M_* = 0.3~\mathrm{M_\odot}$), when the rotational speed drops to close to zero and becomes much smaller than the infall speed. At the early epochs, the rotation remains significant, with a speed comparable to the infall speed. However, there is no clear evidence for a rotationally supported structure either from the column density map or velocity profiles for this moderately weak AD case. The appearance of the circumstellar region at the last epoch is dominated by low-density, expanding DEMS, as in the least diffusive model of M0.0AD0.1.

The appearance of the circumstellar region changes drastically as the ambipolar diffusion coefficient $Q_\mathrm{A}$ increases further to the standard value (i.e., Model M0.0AD1.0). As is seen from the middle column of Fig.~\ref{fig:AD_column_density}, a small dense circumstellar structure is (barely) visible at the earliest epoch ($M_* = 0.1~\mathrm{M_\odot}$). It rotates close to the Keplerian value, and appears to be the seed for the larger, more prominent, rotationally supported structure that develops later. The formation of a rotationally dominated (perhaps even supported) structure of ${\sim} 30~\mathrm{au}$ in radius is evident at the epoch of $M_* = 0.15~\mathrm{M_\odot}$ from the velocity profiles, which show a nearly Keplerian rotation for the structure that clearly dominates the slow (sub-free-fall) infall. This structure is visible in the column density map (see panel h of Fig.~\ref{fig:AD_column_density}) as the small red region (of high column density) near the sink particle (marked by the white cross). A well-defined dense spiral structure is apparent at the epoch when $M_* = 0.2~\mathrm{M_\odot}$. The structure becomes more ring-like at even later epochs ($M_* = 0.25$ and $0.3~\mathrm{M_\odot}$), although arm-like features are still visible. There is little doubt that these dense spiral/ring-like structures are rotationally supported, because they rotate at (or slightly above) the Keplerian value and their infall speed is close to zero (well below the infall value; see the green dash-dotted curves in panels c-e of Fig.~\ref{fig:AD_vel}). Despite the formation of a rotationally supported structure that is absent from the less magnetically diffusive models, Model M0.0AD1.0 retains an important feature of the less diffusive models: the low-density, expanding, DEMS. The co-existence of a rotationally supported disk and DEMS is an interesting new feature that has not been reported in the literature before.

The low-density DEMS all but disappear as the AD coefficient increases further, to 3 and 10 times the standard value (Model M0.0AD3.0 and M0.0AD10.0 respectively). For these more magnetically diffusive cases, the circumstellar region remains dominated by a rotationally supported structure even at late epochs (see the last two columns of Fig.~\ref{fig:AD_column_density}). Compared to Model M0.0AD1.0 with the standard AD coefficient, the rotationally supported structures emerge earlier, are larger at the same epoch (with the same $M_*$), and appear to be more gravitational unstable, as evidenced by the presence of secondary fragments (see, e.g., panels o and s). Although more refined treatments, such as radiation hydrodynamics, are required to investigate the process of gravitational fragmentation properly, the broad trend is unmistakable: as the level of ambipolar diffusion increases, more angular momentum is retained by the circumstellar material, making the formation of a rotationally supported structure easier. In what follows, we seek to understand the physical reasons behind this trend, starting with a discussion of an AD-induced structure that is absent in the ideal MHD limit.

\subsection{AD-induced Diffusion-DEMS}
\label{sec:AD_ideal}

The presence of ambipolar diffusion in principle allows the magnetic flux dragged to the vicinity of the protostar to diffuse outward relative to the protostellar accretion flow, which is a mode of flux redistribution not available in the ideal MHD limit. \citet{1996ApJ...464..373L} showed analytically that the magnetic flux left behind by the accreted stellar material tends to create a circumstellar region of strong magnetic field that is confined by the ram pressure of the accretion flow. The transition between the accretion flow and the circumstellar region dominated by the (AD) redistributed flux is often mediated by a shock of continuous type (C-shock; see \citealp{1993ARA&A..31..373D} for a review) although not always. This transition is a key difference between protostellar accretion with ambipolar diffusion and in the ideal MHD limit. It has been confirmed in non-ideal MHD simulations in 1D \citep[adopting the so-called `thin-disk' approximation, e.g.,][]{2002ApJ...580..987K,2005ApJ...618..783T}, 2D \citep[assuming axisymmetry, e.g.,][]{2009ApJ...698..922M,2010MNRAS.408..322K,2011ApJ...738..180L}, but not yet in 3D. \citet{2012ApJ...757...77K} showed that the AD-induced circumstellar structure found in their 2D (axisymmetric) simulations quickly became unstable when the assumption of axisymmetry is removed. This leaves open the question whether such a structure can ever be produced in 3D in the first place. The answer turns out to be `yes', as we show next.

\begin{figure*}
  \centering
  \includegraphics[width=\textwidth]{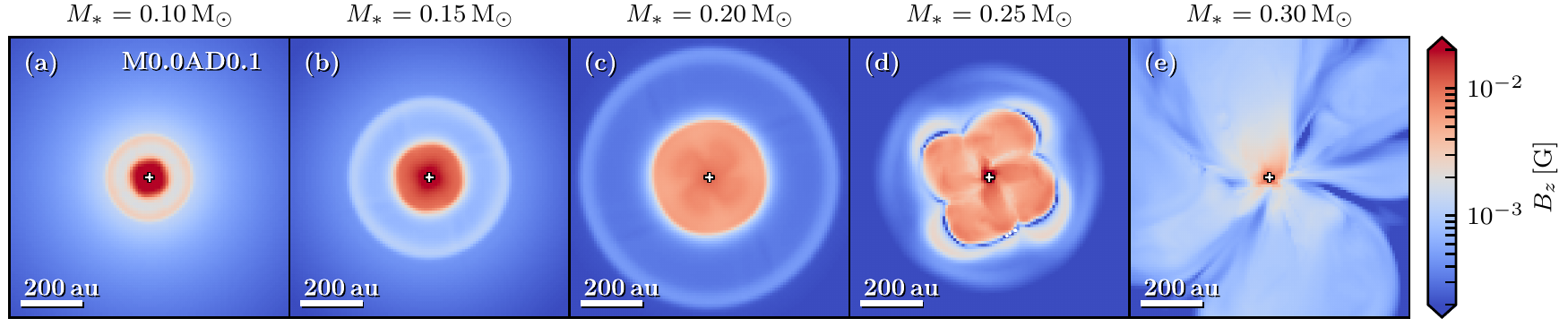}
  \caption{Distribution of the vertical magnetic field on the equatorial plane for the least diffusive model M0.0AD0.1 at the epochs when $M_* = 0.10$, $0.15$, $0.2$, $0.25$, $0.3~\mathrm{M_\odot}$, showing a distinct strong-field plateau at early epochs, which is disrupted at later epochs. (See the supplementary material in the online journal for an animated version of the vertical magnetic field strength and column density distribution.)}
  \label{fig:AD_BzMaps}
\end{figure*}

\begin{figure}
  \includegraphics[width=\columnwidth]{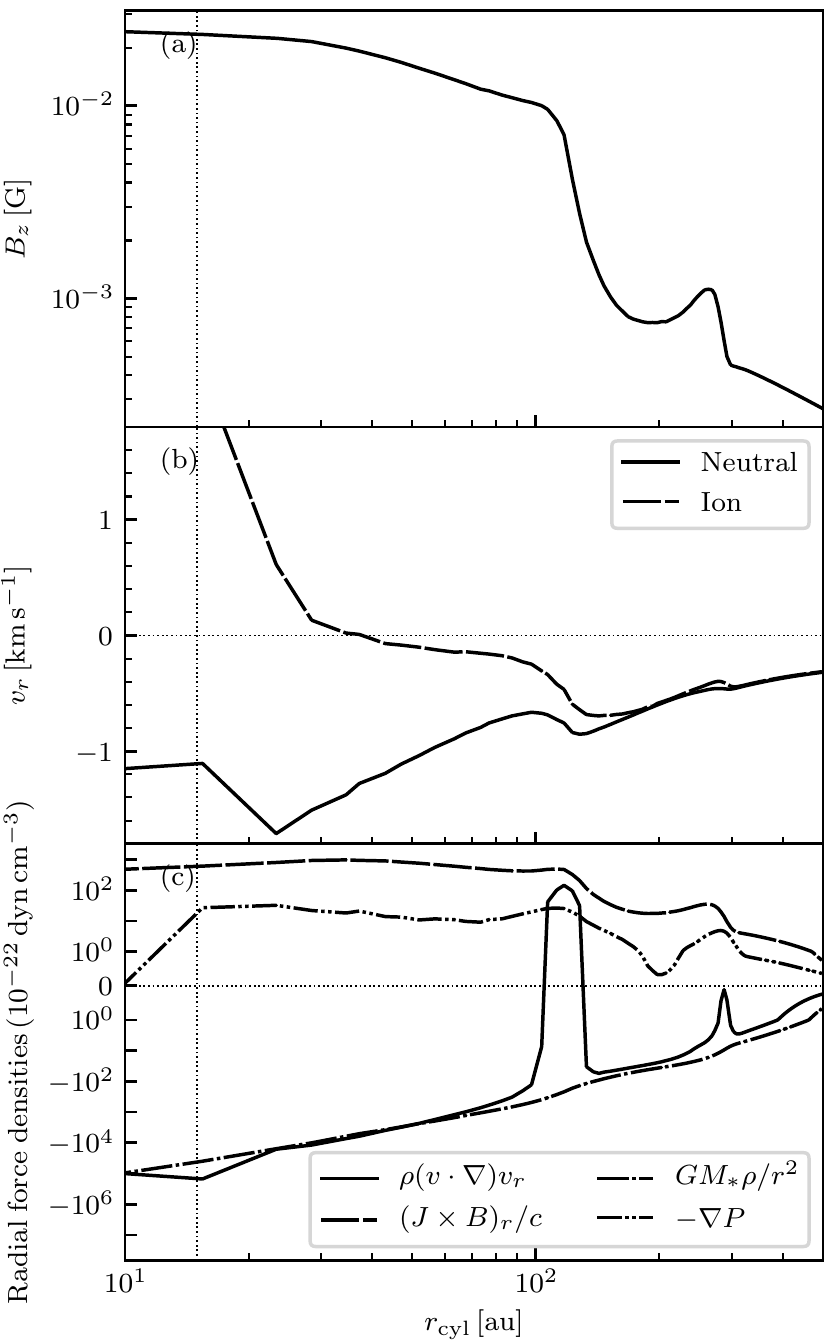}
  \caption{AD-induced magnetically dominated circumstellar structure in Model M0.0AD0.1 at a representative epoch when $M_* = 0.15~\mathrm{M_\odot}$. Plotted are the distributions as a function of radius of (a) the azimuthally averaged vertical magnetic field strength ($B_z$) on the equatorial plane, (b) ion and neutral infall speeds weighted by mass, and (c) the radial component of each of the force terms in the momentum equation (Eq.~\ref{eq:momentum}), including the flow acceleration (the solid line), magnetic force (dashed), gravity of the central object (dash-dotted), and the gas pressure gradient (dash-double-dotted). Note that a symmetrical log scale is used for the vertical axis of panel (c), where the range $[-1, 1]$ is in linear scale to highlight the change of flow acceleration to positive values in the transition region around $r \sim 100~\mathrm{au}$.}
  \label{fig:AD_DEMS}
\end{figure}

%
%
%
The case for the AD-induced structure can be made most clearly in the least diffusive model M0.0AD0.1 at early epochs. There are three lines of evidence supporting this case. First, the model has a clear plateau in the distribution of the vertical magnetic field strength ($B_z$) on the equatorial plane near the center that is distinct from the surrounding (weaker field) region, as seen pictorially in Fig.~\ref{fig:AD_BzMaps}, which plots the maps of $B_z$ at the same 5 epochs as in Fig.~\ref{fig:AD_column_density}. This is further quantified for the representative early epoch $M_* = 0.15~\mathrm{M_\odot}$ in panel (a) of Fig.~\ref{fig:AD_DEMS}, where the azimuthally averaged $B_z$ is plotted as a function of radius. The magnetic flux threading through this plateau region ($r \lesssim 125~\mathrm{au}$, the red region in panel b of Fig.~\ref{fig:AD_BzMaps})
is about $1.30 \times 10^{29}~\mathrm{G\,cm^2}$,
which yields a dimensionless mass-to-flux ratio of 3.7 (using $M_* = 0.15~\mathrm{M_\odot}$ as the mass). 
This ratio is comparable to the mass-to-flux ratio of the central $0.15~\mathrm{M_\odot}$ of the initial core, which is ${\sim} 3.1$ if the core collapse is isotropic and ${\sim} 6.5$ if the collapse is along the field lines. This agreement lends credence to the notion that the plateau is created mostly by the magnetic flux that is decoupled from, and left behind by, the mass already accreted onto the star. The same holds true for the other two early epochs shown in Fig.~\ref{fig:AD_BzMaps}, which correspond to $M_* = 0.1$ and $0.2~\mathrm{M_\odot}$, when the dimensionless ratio of the stellar mass to the magnetic flux in the plateau region is 4.8 and 3.6, respectively.

Second, in the transition zone between the plateau and its surrounding region (at a radius ${\sim} 100~\mathrm{au}$), the infall of the bulk neutral material slows down temporarily, before reaccelerating towards the central protostar (see panel b of Fig.~\ref{fig:AD_DEMS}). The ions, which are tied to the magnetic field lines, collapse much more slowly than the neutrals inside the plateau, however, which is another characteristic of the AD-induced structure proposed originally by \citet{1996ApJ...464..373L} and found numerically in previous 1D and 2D simulations. Since this strongly magnetized structure is created by flux redistribution by (microscopic) magnetic diffusion, we will refer to it as `diffusion-DEMS', to distinguish it from the structure created by interchange instability, where magnetic flux is advected outward by bulk fluid motions of strongly magnetized, low-density regions, which we will sometimes refer to as `advection-DEMS' or just DEMS (since it is the more common of the two types in the current simulations).

Third, at early times when the central stellar mass $M_*$ is of order $0.1~\mathrm{M_\odot}$ or smaller, the magnetic field strength in the diffusion-DEMS agrees to within a factor of 2 with the analytic estimate from \citet{1996ApJ...464..373L} (their equation 7) for the AD-shock based on a balance of the ram pressure of the pre-shock infalling material and the post-shock magnetic pressure. The agreement worsens at later times, when the pre-shock material is confined vertically to a thin layer by the tension force of a highly pinched magnetic field, which is not accounted for in the analytic theory. In such a case, it is more appropriate to exam the balance of forces than pressures.

The mechanics of the accretion flow in the equatorial region is illustrated more clearly in panel (c) of Fig.~\ref{fig:AD_DEMS}, which plots the radial components of all force terms in the momentum equation (Eq.~\ref{eq:momentum}). It is clear that most of the flow acceleration in the radial direction, $\rho \left( \vec{v} \cdot \nabla \right) v_r$ (solid line in the panel), comes from the gravity of the central object (dash-dotted) over most radii.
The main exception is near the plateau-surrounding transition region ($r \sim 100~\mathrm{au}$), where the magnetic force $\left( \vec{J} \times \vec{B} \right)_r/c$ (dashed) and, to a lesser extent, the gas pressure gradient (dash-double-dotted) dominate the gravity and lead to a net outward acceleration, which slows down the accretion flow in the transition region. In other words, the natural tendency for the strong magnetic field in the plateau region to expand is contained by the rapidly collapsing accretion flow.
%
%

A strong-field plateau was also found in the AD simulations of \citet{2015ApJ...801..117T}, \citet{2016A&A...587A..32M}, and \citet{2016ApJ...830L...8H}. \citet{2016ApJ...830L...8H} found an interesting way to interpret the plateau semi-analytically using a one-dimensional model where the inward advection of the magnetic flux is balanced by the outward diffusion of the flux (see their equation~[1]).
The balance is also one of the basic ingredients of our interpretation, based on the semi-analytic work of \citet{1996ApJ...464..373L}, where the neutral material accretes across the magnetic field lines in the diffusion-DEMS (but not necessarily outside this structure), as evidenced by the much slower infall speed of ions (and thus the field lines tied to them) compared to that of neutrals (see panel b of Fig.~\ref{fig:AD_DEMS}). Our interpretation goes one step further and envisions the diffusion-DEMS (or the plateau region) as a region distinct from its surroundings, with a sharp drop in field strength between the two, especially at relatively early epochs (see the first three panels of Fig.~\ref{fig:AD_BzMaps}). 
The drop introduces an outward magnetic force which, in our picture, is reflected in the deceleration of the surrounding accretion flow. This confinement of a lighter fluid (the magnetic field) by a heavier fluid in the presence of the gravity of the central star has long been suspected to be unstable to interchange instabilities \citep[e.g.,][]{1996ApJ...464..373L,2001MNRAS.323..587S}. It is indeed the case, as we show next.

The development of the interchange instability can be seen most clearly in the animations of the distributions of the midplane vertical magnetic field strength ($B_z$) and the surface density $\Sigma$ side-by-side (see online complementary materials or compare the first column of Fig.~\ref{fig:AD_column_density} to that of Fig.~\ref{fig:AD_BzMaps}). The animations show that noticeable azimuthal variations start to develop inside the diffusion-DEMS for both $B_z$ and $\Sigma$ around the epoch when $M_* = 0.2~\mathrm{M_\odot}$, with the two variations anti-correlated.
The variations become more prominent at later epochs, with denser, less magnetized `fingers' infalling towards the central protostar along some azimuthal directions, and less dense but more strongly magnetized pockets expanding away from the central object (see, e.g., panel p of Fig.~\ref{fig:AD_column_density} and panel d Fig.~\ref{fig:AD_BzMaps}), as expected for interchange instability, which is a form of the Rayleigh-Taylor instability. The infalling heavier Rayleigh-Taylor `fingers' deliver both matter and magnetic flux to the central sink region, where the mass is accreted onto the sink particle while the magnetic flux is left behind. Magnetic pressure builds up in the sink region, which is released along the directions of least resistance, driving the expansion of the low-density pockets between the dense infalling `fingers'. Fueled by the magnetic flux released by the accreted mass and the decline of the density of the confining medium, the strongly magnetized, low-density pockets expand quickly at later times, as illustrated in the panel (u) of Fig.~\ref{fig:AD_column_density} and panel (e) of Fig.~\ref{fig:AD_BzMaps} (when $M_* = 0.3~\mathrm{M_\odot}$). Again, the variations of the surface density and $B_z$ are strongly anti-correlated. At such times, the circumstellar region is essentially a mixture of the diffusion-DEMS (driven by the AD-enabled flux redistribution at earlier times) and the advection-DEMS (fueled by the flux released in the sink region at later times)\footnote{We note that the flux decoupling inside the sink region is ultimately achieved through magnetic diffusion as well, so the advection-DEMS is also driven by magnetic diffusion on scales smaller than the DEMS themselves.}.

\subsection{Ambipolar Diffusion and Disk Formation}
\label{sec:AD_diskform}

In this subsection, we will explore how the ambipolar diffusion affects the angular momentum evolution of the protostellar accretion flow compared to the ideal MHD case and how the formation of large disks is enabled by a relatively high ambipolar diffusivity.

\begin{figure}
  \centering
  \includegraphics[width=\columnwidth]{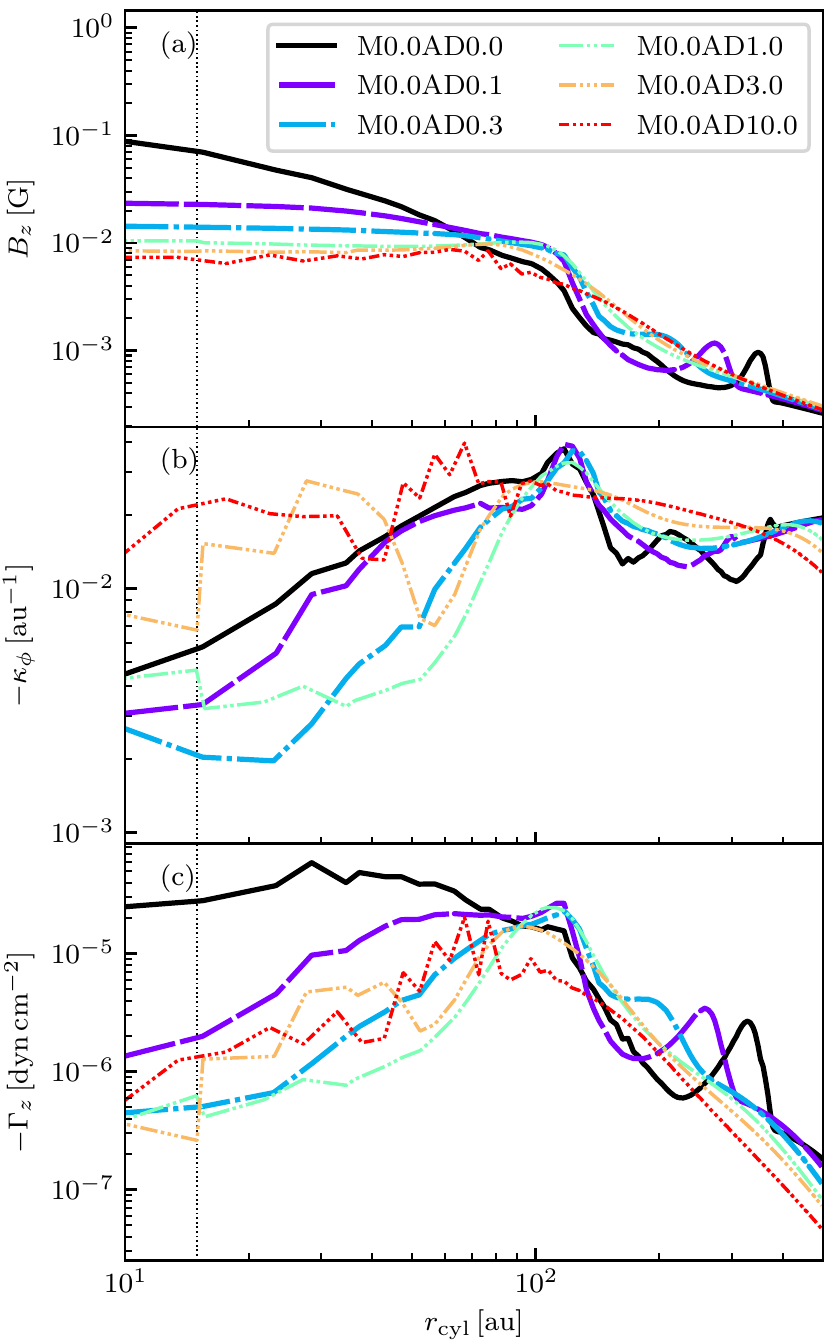}
  \caption{Effects of ambipolar diffusion on the magnetic field strength and structure, and the magnetic braking efficiency. Plotted are the distributions of (a) the azimuthally averaged vertical magnetic field strength $B_z$, (b) the degree of pinching of the magnetic field lines (see equation~\ref{eq:pinching} for a definition) in negative $\phi$-direction, and (c) the magnetic braking torque $-\Gamma_z = (1/c) \left[ \left(\vec{J} \times \vec{B} \right) \times \vec{r} \right]_z$, on the equatorial plane at a representative epoch when $M_* = 0.15~\mathrm{M_\odot}$. The curves in each panel correspond to the models with $Q_\mathrm{A} = 0.0$ (black solid line), $0.1$ (violet dashed), $0.3$ (blue dash-dotted), $1.0$ (green dash-double-dotted), $3.0$ (yellow dash-triple-dotted) and $10.0$ (red dash-quadruple-dotted), respectively.}
  \label{fig:AD_torque}
\end{figure}

We have already seen in \S~\ref{sec:turbulence} that the laminar ideal MHD model M0.0AD0.0 does not form a large rotationally supported structure because of efficient magnetic braking of the protostellar accretion flow. Part of the reason for the efficient braking comes from a rapid increase of the (vertical) magnetic field strength on the equatorial plane ($B_z$) towards the protostar, as illustrated in panel (a) of Fig.~\ref{fig:AD_torque} (black-solid curve). This, coupled with a significant pinching of the field lines in the azimuthal direction (or a strong radial current density $J_r$), gives rise to a large magnetic braking torque ($\propto B_z J_r$) that removes angular momentum from the infalling protostellar envelope efficiently (through a braking-driven outflow).

In the presence of even a weak ambipolar diffusivity, the magnetic field distribution in the circumstellar region is modified significantly, as discussed in the last subsection. Specifically, an ambipolar diffusivity that is 10 times less than the standard value (Model M0.0AD0.1) is enough to limit the increase of $B_z$ towards the protostar, and produce a plateau region of more or less constant $B_z$ -- the diffusion-DEMS. As a result, the protostellar accretion flow is less braked in Model M0.0AD0.1 compared to the ideal MHD model M0.0AD0.0 (compare the black solid and violet dashed curves in panel c of Fig.~\ref{fig:AD_torque}). The weaker braking allows the accretion flow to retain more angular momentum, as can be seen by comparing the violet solid curves in panels a and c of Fig.~\ref{fig:AD_vel} to those in panels a and b of Fig.~\ref{fig:Turb_vel}, respectively.

Despite the reduction in magnetic braking efficiency, this least diffusive model (M0.0AD0.1) does not retain enough angular momentum to form a large rotationally supported structure. As the ambipolar diffusivity increases by a factor of 3 to 0.3 times the standard value (Model M0.0AD0.3), the situation remains qualitatively similar, with the circumstellar region dominated by a well-defined diffusion-DEMS at early epochs and by the development of interchange instabilities at later times (see the second column of Fig.~\ref{fig:AD_column_density}). Quantitatively, the increase in AD coefficient causes a further reduction in the vertical field strength $B_z$ in the circumstellar diffusion-DEMS (compare the blue dash-dotted and violet dashed curves in panel a of Fig.~\ref{fig:AD_torque}). Just as importantly, the circumstellar field lines are significantly less pinched, especially in the azimuthal direction.

To measure the pinching of the field lines quantitatively, we consider the magnetic field curvature:
\begin{equation}
  \vec{\kappa} = \frac{\vec{B}}{B} \cdot \nabla \frac{\vec{B}}{B}.
  \label{eq:pinching}
\end{equation}
In particular, the $\phi$-component of the curvature, $\kappa_\phi$, measures the degree of field line pinching in the azimuthal direction, which is directly tied to magnetic braking. It is significantly smaller for Model M0.0AD0.3 than for Model M0.0AD0.1 within the diffusion-DEMS, as shown in panel (b) of Fig.~\ref{fig:AD_torque} (compare the blue dash-dotted and violet dashed curves). The combination of a weaker and less azimuthally pinched magnetic field reduces the efficiency of magnetic braking in Model M0.0AD0.3 further compared to Model M0.0AD0.1 (compare the blue dash-dotted and violet dashed curves in panel c). However, the reduction is still not enough to enable the formation of a large rotationally supported structure in this relatively weak AD case.

As the AD coefficient increases further from 0.3 to 1.0 times the standard value, a rotationally supported structure is formed, as a result of further weakening of the magnetic braking by a somewhat weaker and substantially less azimuthally pinched circumstellar magnetic field (compare the green dash-double-dotted and blue dash-dotted curves in Fig.~\ref{fig:AD_torque}). The situation with the more magnetically diffusive cases of M0.0AD3.0 and M0.0AD10.0 is qualitatively similar, with the magnetic braking weakened enough to allow for large disk formation. Quantitatively, the rotationally supported structures form at earlier epochs. Note that the magnetic field lines near the protostars are actually more pniched in these two highly diffusive models compared to the standard AD one because of the increased rotational speed (compare, for example, the yellow dash-triple-dotted and green dash-double-dotted curves in panel b of Fig.~\ref{fig:AD_torque}). Nevertheless, the magnetic braking remains weak enough for the large rotationally supported structure to persist until the end of the simulation.
%
%

\section{Disk Formation with Turbulence and Ambipolar Diffusion}
\label{sec:both}

\begin{figure*}
  \centering
  \includegraphics[width=\textwidth]{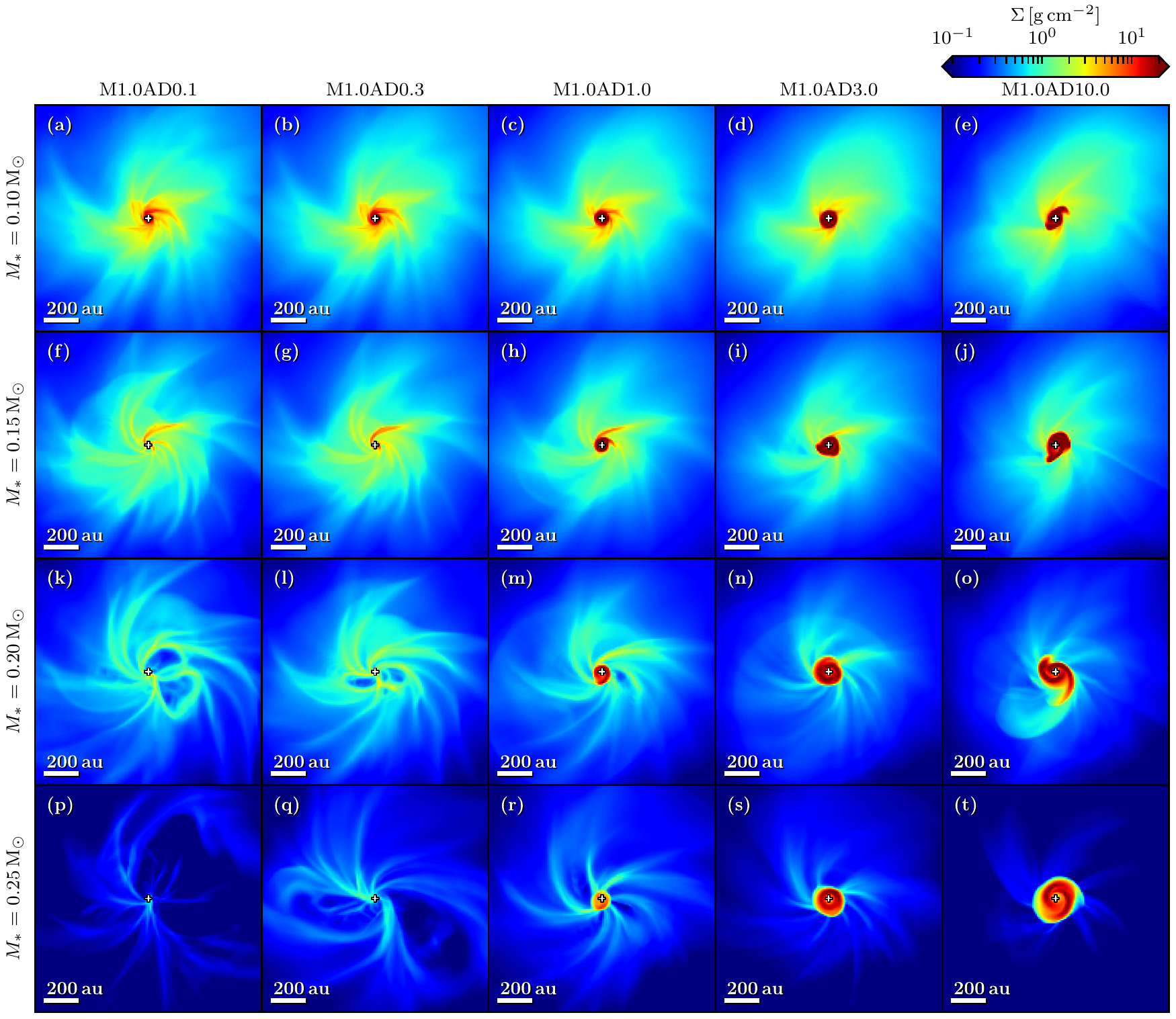}
  \caption{Column density along $z$-axis of the zoom-in simulations of five sonic-turbulent AD models with $Q_\mathrm{A} = 0.1 \times$, $0.3 \times$, $1.0 \times$, $3.0 \times$ and $10.0 \times$ (left to right) the standard value when the sink particle has accreted $0.1$, $0.15$, $0.2$ and $0.25~\mathrm{M_\odot}$ (top to bottom). The sink particle is marked by a cross. (See the supplementary material in the online journal for an animated version of the column density distribution of each model.)}
  \label{fig:AT_column_density}
\end{figure*}

\begin{figure*}
  \centering
  \includegraphics[width=\textwidth]{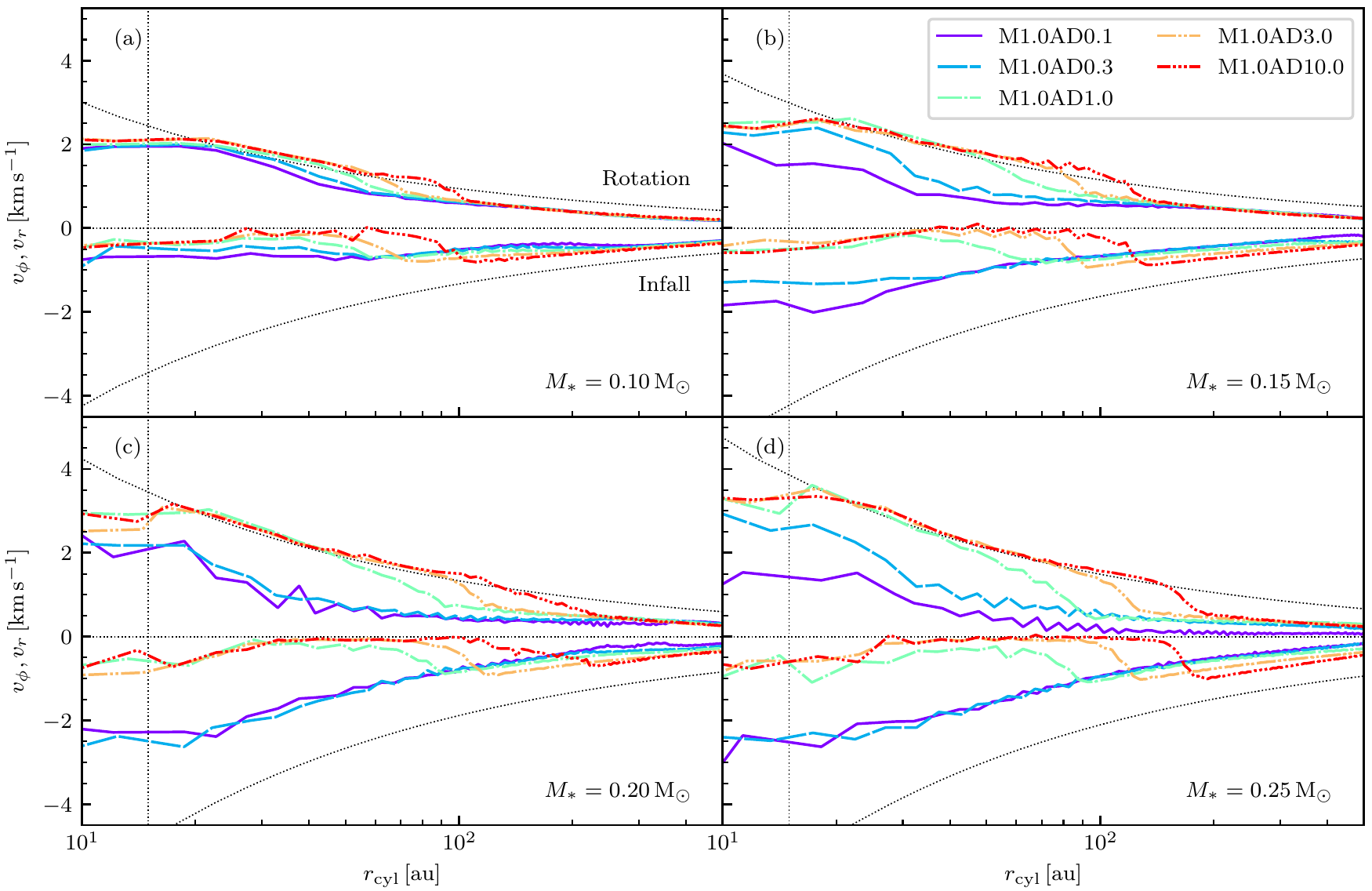}
  \caption{Distributions of the mass-weighted rotational (upper curves) and infall speeds (lower) in a wedge within $45^\circ$ of the equatorial plane compared to the Keplerian (upper black dotted) and free-fall (lower) profile at five representative epochs with stellar mass of $M_* = 0.1$ (panel a), $0.15$ (b), $0.2$ (c) and $0.25~\mathrm{M_\odot}$ (d) for all sonic turbulence AD models with $Q_\mathrm{A} = 0.1 \times$ (violet solid line), $0.3 \times$ (blue dashed), $1.0 \times$ (green dash-dotted), $3.0 \times$ (yellow dash-double-dotted) and $10.0 \times$ (red dash-triple-dotted) the standard value. The vertical dotted line in each panel denotes roughly the radius of the sink region.}
  \label{fig:AT_vel}
\end{figure*}

In the last two sections, we have explored separately the effects of turbulence and ambipolar diffusion on disk formation. Here we study the combined effects of these two physical ingredients, focusing on the cases with a sonic turbulence (with Mach number $\mathcal{M} = 1$) and a range of ambipolar diffusivity, from 0.1 to 10 times the standard value. The results are shown in Fig.~\ref{fig:AT_column_density} and \ref{fig:AT_vel}, which plot, respectively, the surface density and the radial profiles of the mass-weighted infall and rotational speeds at four representative epochs when $M_* = 0.1$, $0.15$, $0.2$ and $0.25~\mathrm{M_\odot}$. It is immediately apparent that well-formed disks are present at the earliest epoch shown ($M_* = 0.1~\mathrm{M_\odot}$) for all five models independent of the values of AD coefficients, both in morphology (the first row of Fig.~\ref{fig:AT_column_density}) and in kinematics (with rotational speed close to the Keplerian value, and much larger than the infall speed; see panel a of Fig.~\ref{fig:AT_vel}). This is very different from the laminar ($\mathcal{M} = 0$) cases where, at the same epoch, a rotationally supported structure either does not exist (see panels a and b of Fig.~\ref{fig:AD_column_density} and the curves plotted in violet solid and blue dashed line in panel a of Fig.~\ref{fig:AD_vel}) or is barely visible (see the right three panels of the same row of Fig.~\ref{fig:AD_column_density} and the rest of the curves in Fig.~\ref{fig:AD_vel}). This is clear evidence that turbulence is beneficial to disk formation, at least at early epochs, independent of the strength of ambipolar diffusion. It is consistent with the results discussed in \S~\ref{sec:turbulence} for the ideal MHD cases.

Whether the rotationally supported structure formed at early epochs can persist to later times or not depends on the value of the AD coefficient. In \S~\ref{sec:turbulence}, we have already seen that the rotationally supported structure formed early in the ideal MHD sonic turbulence model M1.0AD0.0 disappears at later times, with the circumstellar region becoming increasingly dominated by strongly magnetized, low-density (advection-)DEMS. This behavior is preserved qualitatively for the two weakest AD cases (Model M1.0AD0.1 and M1.0AD0.3), where the well-defined rotationally supported disk at the epoch $M_* = 0.1~\mathrm{M_\odot}$ becomes severely disrupted by the epoch $M_* = 0.15~\mathrm{M_\odot}$ and almost completely disappears by the epoch $M_* = 0.2~\mathrm{M_\odot}$ (see the left two columns of Fig.~\ref{fig:AT_column_density} and the purple and blue lines in panels c and d of Fig.~\ref{fig:AT_vel}). This late-time behavior is broadly similar to that of the corresponding laminar AD cases (Model M0.0AD0.1 and M0.0AD0.3), although it is easier to observe the development of magnetic interchange instability in the laminar cases, and their circumstellar regions are dominated by the DEMS to a larger extent at late epochs. For these two weakest AD cases, the sonic turbulence produces only initial transient disks, as in the ideal MHD case.

%
%
To produce a large, persistent, rotationally supported structure, a relatively strong ambipolar diffusion is needed, with or without turbulence. The formation of such a structure in the laminar cases with AD coefficient of 1.0, 3.0 and 10.0 times the standard value has already been discussed in the last section (see the last three columns of Fig.~\ref{fig:AD_column_density}). For these cases of relatively strong ambipolar diffusion, the sonic turbulence makes the small rotationally supported structure at early epochs (e.g., $M_* = 0.1$ and $0.15~\mathrm{M_\odot}$) much more prominent and better defined compared to the laminar cases (contrast the first two panels of the last three columns of Fig.~\ref{fig:AD_column_density} and \ref{fig:AT_column_density}). In addition, the turbulence appears to have made the rotationally supported structure more stable against gravitational fragmentation, judging from the absence of prominent fragments that are prevalent in the corresponding laminar cases at later epochs. Part of the reason is likely that the disk produced in the presence of turbulence is already highly structured (and more strongly magnetized; see \S~\ref{sec:discussion} below) to begin with, which facilitates the redistribution of angular momentum inside the disk and lessens the need for strong spirals to develop gravitationally to transport angular momentum. 
In any case, we have shown that ambipolar diffusion and turbulence work together constructively to form large, persistent, stable disks throughout the protostellar accretion phase, with the turbulence making the disk formation easier at early epochs and ambipolar diffusion making it easier for the disks to survive to later epochs.

\begin{figure}
  \centering
  \includegraphics[width=\columnwidth]{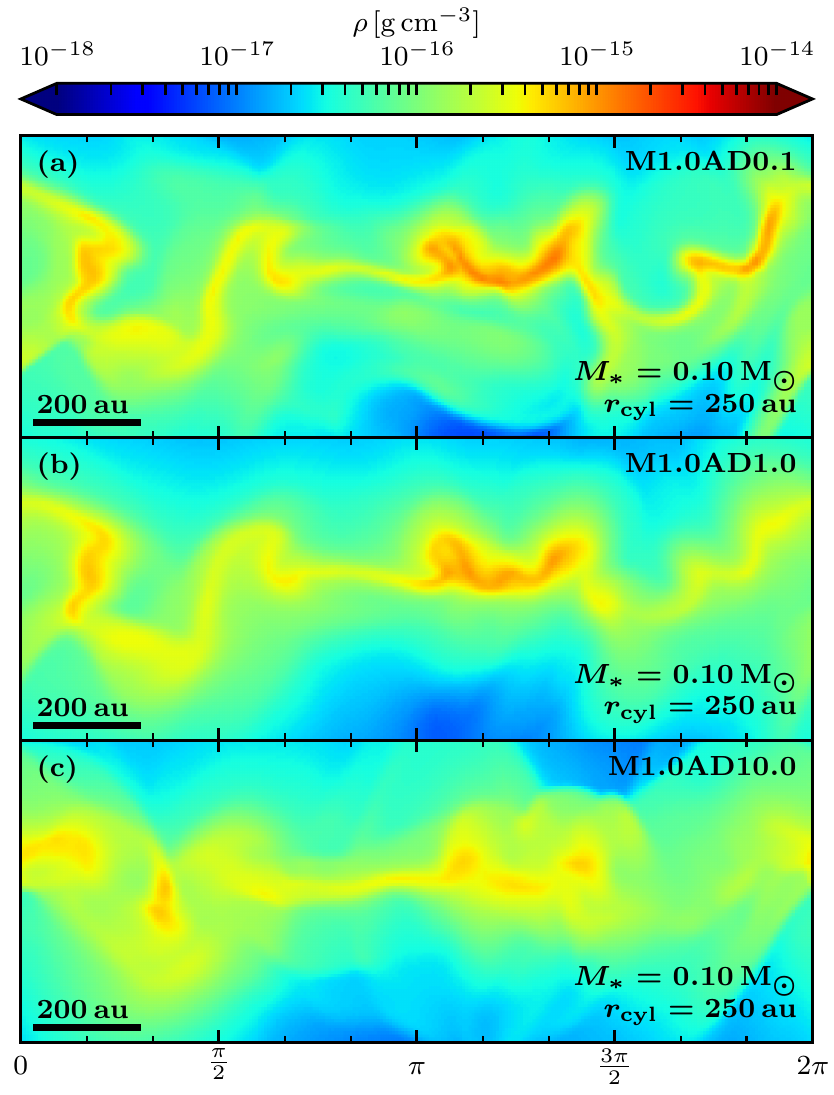}
  \caption{Turbulence-induced pseudodisk warping with ambipolar diffusion. Plotted are the density distributions on a cylinder of radius $r_\mathrm{cyl} = 250~\mathrm{au}$ at the epoch when $M_* = 0.1~\mathrm{M_\odot}$ for Model M1.0AD0.1 (panel a), M1.0AD1.0 (b), and M1.0AD10.0 (c) as a function of azimuthal angle $\phi$ and height $z$, showing a thicker pseudodisk as the AD coeffiecent increases.}
  \label{fig:AT_density_on_cylinder}
\end{figure}

As in the ideal MHD case, the promotion of disk formation at early times by turbulence in the AD cases is facilitated by the warping of pseudodisks. The warp is illustrated in Fig.~\ref{fig:AT_density_on_cylinder}, which plots the density distribution on a cylinder of a representative radius $r_\mathrm{cyl} = 250~\mathrm{au}$ at the epoch $M_* = 0.1~\mathrm{M_\odot}$ for the cases with AD coefficient of 0.1, 1.0 and 10.0 times the standard value. These are to be compared with panel (c) of Fig.~\ref{fig:Turb_density_on_cylinder} for the ideal MHD case. As the AD coefficient increases, the warped pseudodisk appears somewhat thicker, which is understandable since the field lines are expected to be less pinched across the pseudodisk, leaving it less magnetically compressed. Nevertheless, the AD of the levels considered in this paper does not fundamentally change the basic picture of a flattened pseudodisk as the backbone of the protostellar accretion flow and its warping (but not complete disruption) by turbulence. The beneficial effects of pseudodisk warping in disk formation as discussed for the ideal MHD case in \S~\ref{sec:turbulence}, including self-sorting of materials of different specific angular momenta and easier escape of trapped magnetic flux, are therefore preserved.
%
%

\begin{figure*}
  \centering
  \includegraphics[width=\textwidth]{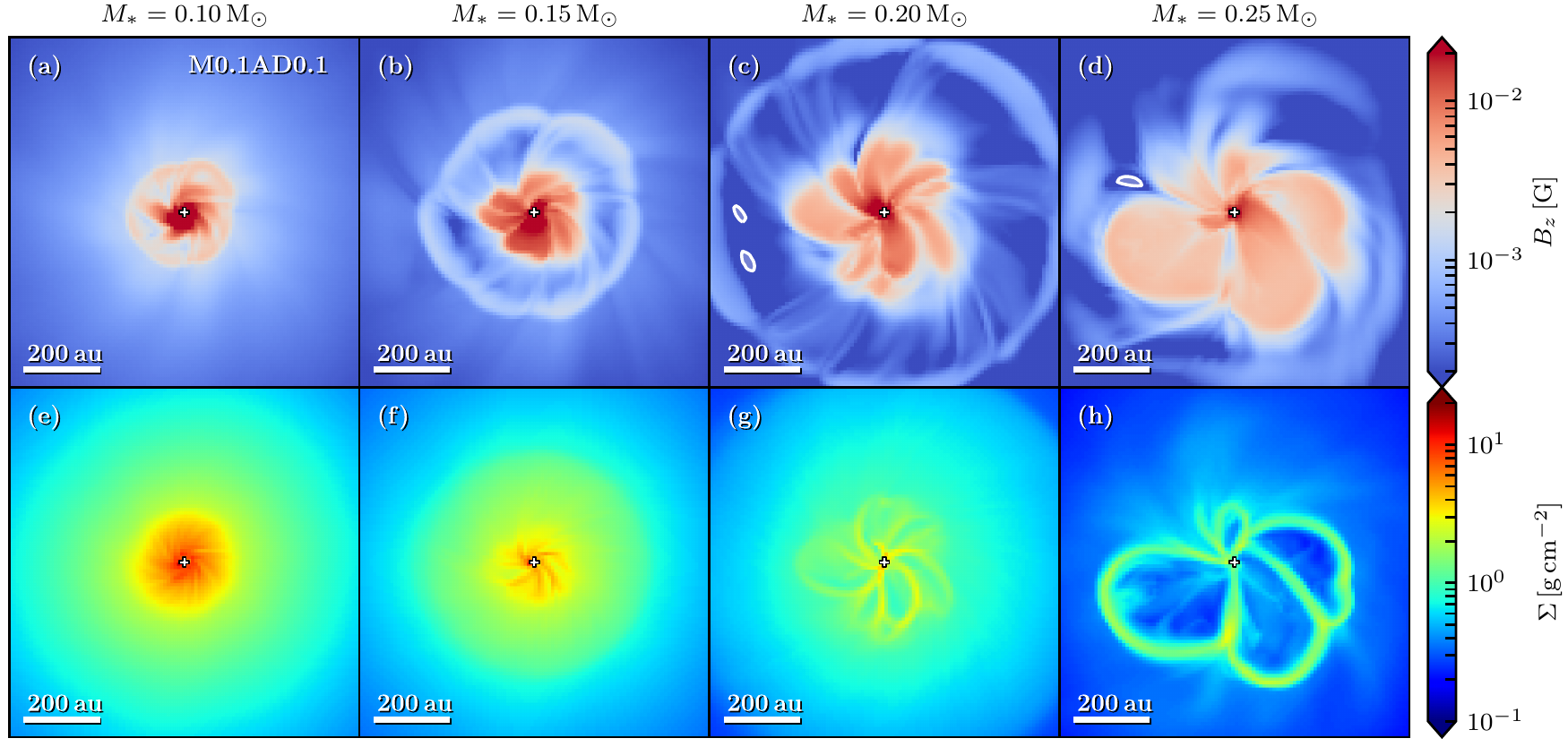}
  \caption{Distribution of the vertical magnetic field on the equatorial plane (top panels) and the column density (bottom panels) for the weak turbulence and weak AD model (M0.1AD0.1) at the epochs when $M_* = 0.05$, $0.1$, $0.15$, $0.2$, $0.25~\mathrm{M_\odot}$, showing a distinct strong-field plateau at early epochs, which is disrupted at later epochs. (See the supplementary material in the online journal for an animated version of the vertical magnetic field strength and column density distribution.)}
  \label{fig:AT_DEMS}
\end{figure*}

The early formation of a large, rotationally supported disk enabled by turbulence is expected to affect the diffusion-DEMS formed in the laminar AD simulations discussed in \S~\ref{sec:AD_ideal}. This interesting structure was particularly well defined in the weakest AD model of M0.0AD0.1 at early epochs (see Fig.~\ref{fig:AD_BzMaps}). In the presence of a relatively weak turbulence of $\mathcal{M} = 0.1$ (Model M0.1AD0.1), the strong-field plateau region (the diffusion-DEMS) at the earliest two epochs is significantly perturbed but not destroyed, as illustrated in the upper row of Fig.~\ref{fig:AT_DEMS}, which plots the distributions of the vertical component of the magnetic field on the equatorial plane ($B_z$) at 4 representative epochs when $M_* = 0.1$, $0.15$, $0.2$, and $0.25~\mathrm{M_\odot}$. In addition, the ring-like structure outside the diffusion-DEMS at intermediate epochs for the laminar case (see panels b-d of Fig.~\ref{fig:AD_BzMaps}) where the vertical field strength is locally enhanced is largely preserved (although significantly perturbed, see panels b-d of Fig.~\ref{fig:AT_DEMS}), again indicating that the weak turbulence does not change the flow structure fundamentally. Nevertheless, it does produce an azimuthal variation that appears to have accelerated the development of the interchange instability which, as in the laminar case, dominates the circumstellar region at late epochs, as shown more clearly in the surface density maps plotted in the lower row of Fig.~\ref{fig:AT_DEMS}. In this case, the combination of a weak turbulence and a weak ambipolar diffusion was not able to enable the formation of a large, well-defined disk. The situation is broadly similar for the somewhat stronger turbulence model of M0.3AD0.1 (with $\mathcal{M} = 0.3$), where the diffusion-DEMS is harder to identify (not shown) and a rotationally supported structure remains absent. As the turbulence level increases to the sonic value ($\mathcal{M} = 1$), the diffusion-DEMS is no longer clearly visible; the circumstellar region at the earliest epoch is dominated by a transient rotationally supported structure instead (see the left column of Fig.~\ref{fig:AT_column_density}).

\section{Discussion}
\label{sec:discussion}

\subsection{How Strongly Magnetized Are Protostellar Disks?} 
\label{sec:2disk}

It is widely believed that a main driver of disk evolution is the magnetic field, through either the magneto-rotational instability \citep{1991ApJ...376..214B} or a magnetic disk-wind \citep{1982MNRAS.199..883B}. How fast the disk evolves depends on the degree of disk magnetization, especially the strength of the poloidal magnetic field threading the disk. For example, in the 2D (axisymmetric) non-ideal MHD simulations of \citet{2017ApJ...845...75B}, a mid-plane plasma-$\beta$ of $10^5$ for the initial poloidal magnetic field is enough to drive an accretion rate of order $10^{-8}~\mathrm{M_\odot\,yr^{-1}}$, typical of classical T Tauri stars. The rate is increased to ${\sim} 10^{-7}~\mathrm{M_\odot\,yr^{-1}}$ when the initial poloidal field strength is increased by a factor of $\sqrt{10}$, corresponding to a plasma-$\beta$ of $10^4$. The degree of disk magnetization is currently a critical free parameter in modeling the dynamics of protoplanetary disks that is unfortunately difficult to measure observationally. Theoretically, it is expected to be determined by the amount of magnetic flux carried into the disk during its formation out of the dense molecular cloud cores, which are known to be significantly magnetized, and the subsequent magnetic flux transport within the disk \citep[e.g.,][]{2014ApJ...785..127O}. Determining the initial degree of magnetization of protostellar disks is therefore a central task of magnetized disk formation calculations. In this subsection, we will take a first step in this direction.

\begin{figure*}
  \centering
  \includegraphics[width=\textwidth]{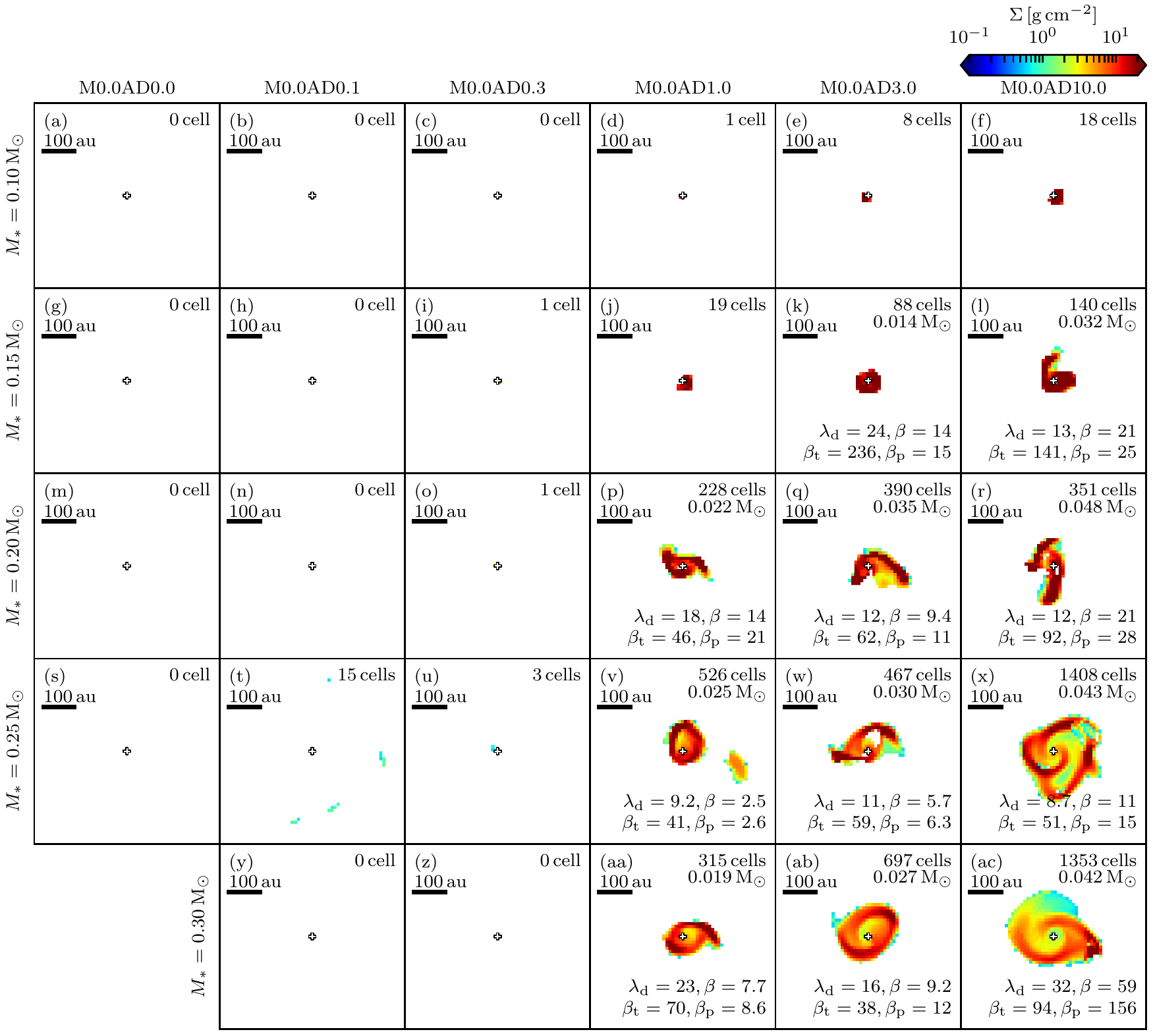}
  \caption{Column density of cells identified as a part of the disk along $z$-axis of all non-turbulent AD models with $Q_\mathrm{A} = 0.0 \times$, $0.1 \times$, $0.3 \times$, $1.0 \times$, $3.0 \times$ and $10.0 \times$ (left to right) the standard value when the sink particle has accreted $0.1$, $0.15$, $0.2$, $0.25$ and $0.3~\mathrm{M_\odot}$ (top to bottom). The sink particle is marked by a cross. The number of cells is written on the upper right corner. For those snapshots where either underdeveloped  or well-developed disks (with more than 50 disk cells) are identified, the disk mass is shown below the disk cell number in the upper right corner, and the mass-to-flux ratio $\lambda_\mathrm{d}$, total plasma-$\beta$, and toroidal component $\beta_\mathrm{t}$ and polaroidal component $\beta_\mathrm{p}$ of the plamsa-$\beta$ are shown on the lower part of each panel.}
  \label{fig:AD_disk}
\end{figure*}

\begin{figure*}
  \centering
  \includegraphics[width=\textwidth]{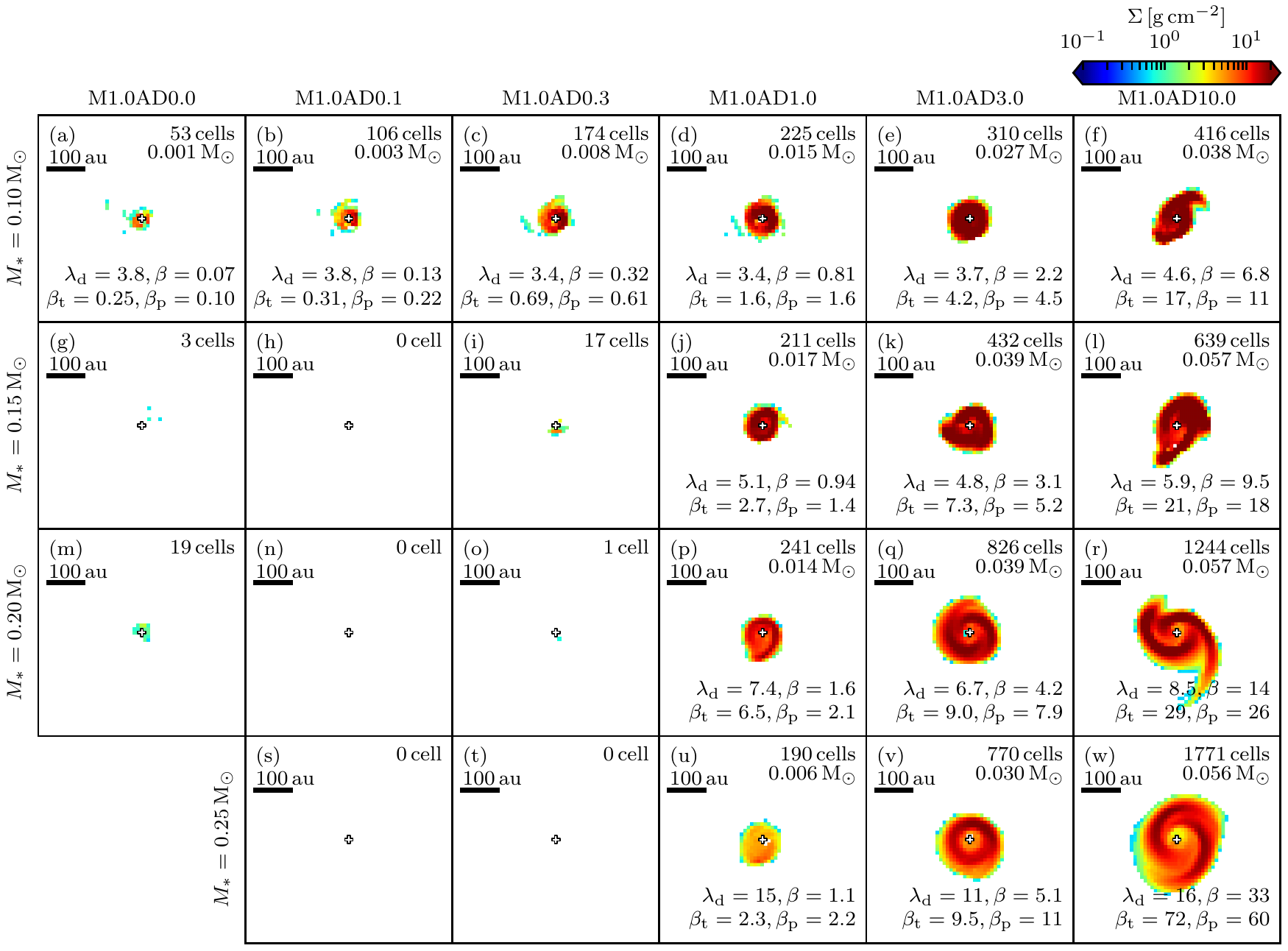}
  \caption{Column density of cells identified as a part of the disk along $z$-axis of all sonic-turbulent AD models with $Q_\mathrm{A} = 0.0 \times$, $0.1 \times$, $0.3 \times$, $1.0 \times$, $3.0 \times$ and $10.0 \times$ (left to right) the standard value when the sink particle has accreted $0.1$, $0.15$, $0.2$ and $0.25~\mathrm{M_\odot}$ (top to bottom). The sink particle is marked by a cross. The number of cells is written on the upper right corner. For those snapshots where either underdeveloped disks or well-developed disks are identified, the disk mass is written on the upper right corner, and the mass-to-flux ratio $\lambda_\mathrm{d}$, total plasma-$\beta$, and toroidal component $\beta_\mathrm{t}$ and polaroidal component $\beta_\mathrm{p}$ of $\beta$ are written on the lower right corner of each panel.}
  \label{fig:AT_disk}
\end{figure*}

To quantify the disk properties, particularly its degree of magnetization, we need to determine which simulation cells belong to the disk. We will adopt the criteria used in \citet{2016A&A...587A..32M}, which are:
\begin{enumerate}
  \item the material in the cell is close to the hydrostatic equilibrium in the $z$-direction so that the rotational speed is significantly greater than the vertical speed ($v_\phi > f \lvert v_z \rvert$);
  \item it is rotationally supported against infall so that the rotational speed is significantly greater than the radial speed ($v_\phi > f \lvert v_r \rvert$);
  \item it is significantly dominated by rotational support instead of thermal support ($\rho v_\phi^2 / 2 > f P$);
  \item it has high density ($\rho > 3.8 \times 10^{-15}~\mathrm{g\,cm^{-3}}$).
\end{enumerate}
Following \citet[see also \citealp{2018MNRAS.473.2124G}]{2016A&A...587A..32M}, we will choose a value of 2 for the factor $f$, which ensures that the quantities to be compared are significantly different. The results are shown in Fig.~\ref{fig:AD_disk} and \ref{fig:AT_disk}.

In Fig.~\ref{fig:AD_disk}, we plot the column density along the $z$-axis for the `disk cells' selected based on the above criteria for all non-turbulent ($\mathcal{M} = 0$) models that have a wide range of values for the AD coefficient, from $Q_\mathrm{A}=0$ (ideal MHD, the leftmost column) to $10$ (rightmost column), at five different epochs (the last epoch was not reached in the ideal MHD case where the mass in the simulation box is significantly depleted by a magnetic braking driven outflow). Also shown for each model and each epoch are the number of disk cells and the total mass in these cells. Obviously as the number of disk cells increases, the disk becomes better defined, although the boundary between a disk and not a disk is somewhat arbitrary. For definiteness, we will refer to a structure with less than 50, between 50 and 100, and more than 100 disk cells as `no disk', `underdeveloped disk', and `well-developed disk' (or `disk' for short), respectively. Based on this definition, we find no discernible disk at any epoch for the laminar ideal MHD (Model M0.0AD0.0) and laminar, relatively weak AD (M0.0AD0.1 and M0.0AD0.3) models (see the first three columns in Fig.~\ref{fig:AD_disk}).

A well-developed disk does appear in the standard AD coefficient case (M0.0AD1.0), at the epoch when $M_* = 0.2~\mathrm{M_\odot}$ (panel p) and later. For such well-developed disks, we characterize their degree of magnetization through two dimensionless numbers: the normalized ratio of the disk mass to the magnetic flux threading the disk, $\lambda_\mathrm{d}$, and the plasma-$\beta$ (the ratio of thermal to magnetic pressure), both noted in Fig.~\ref{fig:AD_disk}. The former is to be compared to the mass-to-flux ratio of the initial cloud core, which is $\lambda_\mathrm{core} = 2.6$ globally. For the standard AD case under discussion, we find $\lambda_\mathrm{d} = 18$ at the epoch when $M_* = 0.2~\mathrm{M_\odot}$. It is much higher than $\lambda_\mathrm{core}$, indicating that the newly formed disk is much less magnetized relative to its mass compared to its parental core, presumably because of the action of ambipolar diffusion, which is expected to redistribute the magnetic flux outward, away from the high-density circumstellar region. Nevertheless, this disk is still significantly magnetized, as reflected in the value of the plasma-$\beta$, which is 14 at this epoch. This value, while much larger than that for the initial core as a whole ($\beta_\mathrm{core} = 1.9$), is much smaller than what is typically adopted in MHD simulations of protoplanetary disks, as mentioned earlier. Interestingly, the magnetic energy is dominated by the poloidal component of the magnetic field rather than the toroidal component, as shown by the values of $\beta_\mathrm{t}$ and $\beta_\mathrm{p}$ in Fig.~\ref{fig:AD_disk} for each disk, which are the ratios of the thermal to magnetic energy due to the toroidal and poloidal field component respectively. The disk remains significantly magnetized at later epochs. The values of $\lambda_\mathrm{d}$ and $\beta$ are formally $9.2$ and $2.5$, respectively, at the epoch when $M_* = 0.25~\mathrm{M_\odot}$, smaller than their counterparts at the earlier epoch of $M_* = 0.2~\mathrm{M_\odot}$, but they are affected by a relatively low density region that satisfies the disk criteria but is detached from the main body of the disk (see panel v of Fig.~\ref{fig:AD_disk}). By the last epoch shown (panel aa, $M_* = 0.3~\mathrm{M_\odot}$), the disk has $\lambda_\mathrm{d} = 23$ and $\beta = 7.7$, with the magnetic energy strongly dominated by the poloidal field component.

As the AD coefficient increases, the disk starts to form earlier. For example, by the epoch when $M_* = 0.15~\mathrm{M_\odot}$, a disk is already well-developed in the most magnetically diffusive model (M0.0AD10.0, panel l), clearly visible although under-developed in the second most diffusive model (M0.0AD3.0, panel k), but barely discernible in the standard AD case (panel j). Another trend is that, as the AD coefficient increases, the disk is somewhat bigger (compare the disks in the last three columns of Fig.~\ref{fig:AD_disk}) and more massive, with the disk mass increasing from ${\sim} 0.02~\mathrm{M_\odot}$ for Model M0.0AD1.0 to ${\sim} 0.03~\mathrm{M_\odot}$ for M0.0AD3.0 to ${\sim} 0.04~\mathrm{M_\odot}$ for M0.0AD10.0. The disk remains significantly magnetized, with $\lambda_\mathrm{d} \sim 10-20$ and $\beta \sim 10-20$ typically, and the magnetic energy dominated by the poloidal field component in most cases. The main exception is the last epoch of the most diffusive model (M0.0AD10.0, $M_* = 0.3~\mathrm{M_\odot}$), when the disk is significantly less magnetized, with $\lambda_\mathrm{d} = 32$, $\beta = 59$, and the magnetic energy dominated by the toroidal field component instead.

Turbulence changes the formation and properties of disks substantially, as illustrated by Fig.~\ref{fig:AT_disk}, which is the same as Fig.~\ref{fig:AD_disk} but for all models with sonic turbulence ($\mathcal{M} = 1$) and different levels of ambipolar diffusion. The change is the most pronounced at the earliest epoch ($M_* = 0.1~\mathrm{M_\odot}$), when the disk is well developed in all AD models and is clearly visible (although under-developed) even for the ideal MHD case (panel a). The ideal MHD disk is strongly magnetized, with $\lambda_\mathrm{d} = 3.8$ that is comparable to that of its parental core (especially the central part that is less magnetized compared to the core as a whole) and a plasma-$\beta$ that is much less than unity ($\beta_\mathrm{d} = 0.1$). The strong magnetization, we believe, is the reason that the disk enabled by turbulence early in the ideal MHD case is transient; the magnetic braking is simply too efficient to allow the disk to persist for a long time. The same is broadly true for the early disks formed in the two weakest AD cases (Model M1.0AD0.1 and M1.0AD0.3), even though their masses ($0.003$ and $0.008~\mathrm{M_\odot}$) are significantly larger than that of the ideal MHD disk ($0.001~\mathrm{M_\odot}$); both disks disappear at later epochs.

The early disk enabled by turbulence does survive to the last epoch shown in Fig.~\ref{fig:AT_disk} ($M_* = 0.3~\mathrm{M_\odot}$) in the standard AD model of M1.0AD1.0. However, even in this case where a disk is formed without turbulence, the properties of the disk, especially its magnetization, are strongly affected by the turbulence. Similar to the ideal MHD and the two weakest AD cases, the early disk is strongly magnetized, with $\beta = 0.81$ and $0.94$ at the epochs $M_* = 0.1$ and $0.15~\mathrm{M_\odot}$ respectively (see panels d and j). The order-of-unity disk plasma-$\beta$ persists to the later epoch of $M_* = 0.2~\mathrm{M_\odot}$, when $\beta = 1.6$. This value is much smaller than that for the corresponding non-turbulent model at the same epoch, where $\beta = 14$. The stronger magnetization induced by turbulence is expected to make it more difficult for the disk to survive. This is indeed true for the standard AD case, where the disk mass is reduced by more than a factor of 2 (from $0.014$ to $0.006~\mathrm{M_\odot}$) between the last two epochs ($M_* = 0.2$ and $0.25~\mathrm{M_\odot}$). Indeed, the disk disappears completely at even later epochs (not shown). In this particular case, the turbulence not only enabled the earlier formation of the disk, but also made the disk more strongly magnetized and thus harder to survive. This negative effect of turbulence on the long-term survivability of AD-enabled disks is a new phenomenon that has not been seen before.

The reduction of disk plasma-$\beta$ by turbulence is also evident in the stronger AD models of M1.0AD3.0 and M1.0AD10.0. For example, at the epoch when $M_* = 0.2~\mathrm{M_\odot}$, the well-developed disks in the laminar Model M0.0AD3.0 and M0.0AD10.0 have $\beta = 9.4$ and $21$, respectively. When a sonic turbulence is present, these values are reduced by roughly a factor of 2, to 4.2 for Model M1.0AD3.0 and 14 for Model M1.0AD10.0, respectively. Nevertheless, these significantly magnetized disks are able to survive to the end of the simulation, unlike the standard AD case.

Another interesting effect is that, for the more magnetically diffusive cases where disk formation can be enabled by AD alone, the turbulence makes the disk more stable to violent gravitational instability (which often produces fragments distinct from the main body of the disk in the laminar cases, see, e.g., panels s and o of Fig.~\ref{fig:AD_column_density}). As speculated earlier, part of the reason may be that the turbulence strongly warps the pseudodisk that feeds the disk, leading to a strong initial inhomogeneity in the disk (including spirals) that facilitates the redistribution of angular momentum which, in turn, lessens the need for violent gravitational instability to do so. Another reason is that the disks formed in the presence of turbulence tend to be more strongly magnetized and thus less prone to gravitational instability. In any case, the combination of a relatively strong turbulence and relatively strong AD appears capable of producing a large, persistent, stable, but significantly magnetized disk.

The significant level of magnetization that we found in the disk formed in the presence of AD and especially turbulence may be a potential problem for the disk evolution in late phases of star formation, particularly the classical T Tauri phase that is generally thought to be crucial to planet formation. While it is reassuring that the disk can inherit a magnetic field from its parental core, our simulations show that the inherited field may be too strong for the protoplanetary disks, with a typical plasma-$\beta \sim 10-20$ without turbulence and ${\sim} 1-10$ with turbulence, and a poloidal field component comparable to, and often larger than, the toroidal field component. These values are orders of magnitude lower than the typical initial values used in global protoplanetary disk simulations (${\sim} 10^4-10^5$), as mentioned earlier. One empirical constraint on the disk plasma$-\beta$ comes from the measurement of the magnetic field strength of ${\sim} 0.54~\mathrm{G}$ in the Semarkona meteorite \citep{2014Sci...346.1089F}. If we adopt the estimates of the temperature and density distributions for the solar nebula from \citet{2007ApJ...671..878D} and assume that the parent body of the meteorite comes from the main asteroid belt at a radius of ${\sim} 2.5~\mathrm{au}$, we obtain a plasma-$\beta$ of ${\sim} 700$ in the gas surrounding the meteorite if the field strength in the gas is the same as that in the meteorite. It is, however, possible that the former may be weaker than the latter, by a factor between 1 and 10, if the chondrules are formed in nebular shocks \citep{2002M&PS...37..183D}. In this case, the gas plasma-$\beta$ would be in the range of ${\sim} 10^3-10^5$, close to the values often adopted in simulations of relatively evolved, protoplanetary disks. This is reassuring since the Semarkona chondrules are thought to form in a rather late stage of the solar nebula evolution, with an inferred age of 2-3 Myrs after the formation of the first calcium alunimum-rich inclusions \citep{2002M&PS...37..421M}. This is much older than the disks studied in this paper, which typically have an age of only a few times $10^4$ years.

The above result needs to be tested with high resolution simulations that resolve the disks better. If confirmed, the discrepancy would indicate that, although ambipolar diffusion and turbulence can enable disks to form, they may not be able to demagnetize the formed disk enough to satisfy the constraints imposed by the (low) accretion rate and measurements of meteoritic magnetic field strength in late phases. Additional magnetic diffusivities, such as Ohmic dissipation and Hall effect, a detailed treatment of the disk thermodynamics, and longer-term simulations of the disk from its initial formation to the T Tauri phase may be required to resolve the discrepancy.


\subsection{Connection with Previous Work and Future Refinement}
\label{sec:refinement}

As discussed in the introduction, the most detailed non-ideal MHD studies of magnetized disk formation to date tend to focus on the early phase up to, and slightly beyond, the formation of Larson's second core (stellar seed; see recent reviews by \citealp{2016PASA...33...10T} and \citealp{2018FrASS...5...39W}). The focus of this work is on the less well explored protostellar accretion phase, with emphasis on turbulence and ambipolar diffusion, which have previously been investigated separately but not in combination thus far.

Our self-gravitating ideal MHD simulations with turbulence (but not AD) can be viewed as an extension of the work by \citet{2014ApJ...793..130L}, who studied the simplified problem of the accretion of non-self-gravitating turbulent rotating protostellar envelope onto a star of fixed mass. Our more self-consistent treatment strengthened their general conclusions that the structure of the magnetized protostellar accretion flow is dominated by a turbulence-warped but spatially coherent pseudodisk, and that the turbulence is beneficial to disk formation. However, there is an important difference: whereas the disk induced by a sonic turbulence in \citet{2014ApJ...793..130L} lasted until the end of their simulation \citep[see also][]{2016ApJ...819...96G}, that in the corresponding model here (Model M1.0AD0.0) is more transient and becomes disrupted by DEMS at late epochs (see the right column of Fig.~\ref{fig:Turb_column_density}). The exact reason for this difference is unclear. We also note that the lack of a persistent rotationally supported disk is broadly consistent with the work of \citet{2018MNRAS.473.2124G}, who found that such a disk does not form unless there is a large misalignment between the turbulence-induced angular momentum (which is set to zero for our simulation as a whole) and the magnetic field. We note that \citet{2017ApJ...846....7K} found multiple spirals in the circumstellar regions in some of their ideal MHD disk formation simulations (see the left column of their Fig.~10) similar to our ideal MHD turbulent models (M0.5AD0.0 and M1.0AD0.0). It would be interesting to see whether their spirals are part of a warped but spatially coherent pseudodisk (as true for our cases, e.g., Fig.~\ref{fig:Turb_3d_pseudodisk}) or not.
%
%
%
%

Our 3D laminar (non-turbulent) simulations with AD is a natural extension of the 2D (axisymmetric) work of \citet{2011ApJ...738..180L}. In particular, we have shown for the first time that the strongly magnetized circumstellar structure driven by AD-induced magnetic flux redistribution that is predicted analytically and found in the 2D simulations can be produced in 3D as well. We have shown further that such a structure is unstable to the magnetic interchange instability, which is broadly consistent with the 3D work of \citet{2012ApJ...757...77K}. However, an apparent difference is that there are AD-enabled, persistent rotationally supported circumstellar structures in our simulations (see Fig.~\ref{fig:AD_column_density}) but not in theirs. This is probably because \citeauthor{2012ApJ...757...77K} adopted an initially uniform density distribution, which is known to be less conducive to disk formation compared to the centrally condensed distribution adopted in most of our simulations \citep[see, e.g.,][]{2014MNRAS.438.2278M}. Indeed, we have performed two simulations with an initially uniform density distribution (keeping the global mass-to-flux ratio the same as before, i.e., $\lambda_\mathrm{core} \approx 2.6$) for the standard AD case with and without a sonic turbulence (Models M1.0AD1.0US and M0.0AD1.0US respectively). Neither model produced a large, persistent rotationally supported circumstellar structure. This is consistent with \citet{2016MNRAS.460.2050Z}, who found that enhanced ambipolar diffusion, possibly by removal of small grains, is needed to enable the disk formation in such cases. It is also broadly consistent with \citet[see also \citealp{2016ApJ...830L...8H}]{2016A&A...587A..32M}, who found that, for an initially uniform core with aligned magnetic field and rotation axis, the formation of a large, well-resolved disk is enabled by AD in the relatively weak field case of $\lambda \approx 4$ (after correcting for a somewhat different definition of the dimensionless mass-to-flux ratio) but barely in the stronger field case of $\lambda \approx 1.6$. They showed further that disk formation in the latter case is helped by a relatively large misalignment (of $40^\circ$) between the magnetic field and rotation axis. We note that our 3D results are also qualitatively consistent with those from 2D (axisymmetric) simulations of \citet{2009ApJ...698..922M} in that relatively large disks can form if the ambipolar diffusion is strong enough. It is, however, difficult to quantitatively compare these early 2D simulations with the current 3D simulations because they have very different initial conditions (self-similar magnetized toroids with non-uniform magnetic fields vs centrally condensed cores threaded by an uniform magnetic field).

The most unique aspect of our work is the combination of turbulence and ambipolar diffusion. We have shown that the two work together to form disks that are much better defined than those induced by turbulence or AD separately. The turbulence tends to promote disk formation at early epochs while AD helps the early disks to survive to later times. However, even with both turbulence and ambipolar diffusion, there is no guarantee that a large, persistent disk would form automatically. It depends on many factors, including the degree of magnetization (i.e., $\lambda_\mathrm{core}$), the level of turbulence (i.e., Mach number $\mathcal{M}$), the structure of the core (e.g., the initial density distribution), the rate of core rotation, and the degree of coupling between the magnetic field and the bulk neutral cloud material. For example, we do not find a large, persistent disk for the uniform core model M1.0AD1.0US with a sonic turbulence and the standard AD coefficient, as mentioned above.

We caution the reader that small, numerically unresolved disks may still form in those simulations that do not produce large, persistent rotationally supported disks. This is a general concern for disk formation simulations that focus on the protostellar mass accretion phase where a sink particle treatment is needed, particularly for the relatively low-resolution simulations presented in this manuscript, because the angular momentum of the material accreted onto the central star (sink particle) from the sink region is lost as far as the disk formation is concerned \citep[e.g.][]{2014MNRAS.438.2278M,2018MNRAS.473.2124G}.
We have carried out a crude resolution study, by running the uniform-grid simulations well into the protostellar accretion phase without zoom-in (256 cells in 5000~au). Compared with the zoom-in simulations presented in this paper, it is somewhat more difficult to form disks in these lower resolution simulations, although the general trend is the same, namely, disks are formed more easily in the presence of a larger ambipolar diffusion and a stronger turbulence.
We plan to perform higher resolution simulations in the near future using a version of the \textsmaller{Athena++} code (currently under development) that will include not only non-ideal MHD effects and sink particles but also a self-gravity solver that works with AMR (adaptive mesh refinement).

%
%
%

\section{Conclusion}
\label{sec:conclusion}

We have carried out a set of numerical simulations of disk formation in rotating, magnetized molecular cloud cores including turbulence and ambipolar diffusion, both separately and in combination, with a focus on the protostellar mass accretion phase of star formation that is made possible by a sink particle treatment. The main results are as follows.

\begin{enumerate}
  \item In the ideal MHD limit, a relatively strong, sonic turbulence on the core scale strongly warps but does not completely disrupt the well-known magnetically-induced flattened pseudodisk that dominates the inner protostellar accretion flow in the laminar case, in agreement with previous results obtained in the absence of self-gravity. The turbulence facilitates the disk formation at early times, possibly by creating strong inhomogeneities (including low-density regions) in the warped pseudodisk that allow the magnetic flux trapped near the forming star to escape more easily. The warping of the pseudodisk may also promote disk formation by making it easier for materials with high specific angular momenta to retain their angular momenta. However, the turbulence-enabled initial disk is too strongly magnetized to persist to the end of the simulation, when the majority of the core material has been accreted onto the star or ejected in outflows. It is replaced at later times by strongly magnetized, low-density expanding regions where the magnetic flux associated with the accreted stellar material is advected outwards (i.e., the so-called `advection-DEMS' or simply DEMS).

  \item We find from our 3D laminar (non-turbulent) non-ideal MHD simulations that ambipolar diffusion can redistribute the magnetic flux associated with the accreted stellar material to a circumstellar region where it is trapped by the surrounding protostellar accretion flow, forming the so-called `diffusion-DEMS', in agreement with previous analytic work and 2D (axisymmetric) simulations. For a relatively weak ambipolar diffusion, the strongly magnetized diffusion-DEMS dominates the circumstellar region at early times, making disk formation difficult. It is subsequently disrupted by the magnetic interchange instabilities, although the circumstellar region remains strongly magnetized, with no evidence for a large rotationally supported disk. As the level of ambipolar diffusion increases, the magnetic field in the circumstellar region becomes weaker and less pinched in the azimuthal direction, both of which reduce the magnetic braking torque, making it easier to form a large persistent disk.

  \item We find from our non-ideal MHD simulations with sonic turbulence that the turbulence and ambipolar diffusion promote disk formation in a complementary manner, with the former ensuring the formation of a relatively large disk early in the protostellar accretion phase and the latter facilitating the survival of the disk to later times. In addition, the turbulence tends to make the disks formed in the presence of a relatively strong ambipolar diffusion more stable to gravitational fragmentation.

  \item The turbulence-enabled early disks tend to be strongly magnetized, which makes them difficult to persist unless a relatively strong ambipolar diffusion is also present. Even with a strong ambipolar diffusion, the disks formed in our simulations remain rather strongly magnetized, with a typical plasma-$\beta$ of order a few tens or smaller, which is 2-3 orders of magnitude lower than the values commonly adopted in MHD simulations of the relatively evolved, slowly accreting, protoplanetary disks. This potential tension highlights the strong need to quantify the evolution of the disk magnetic field from its parental core to the end of its evolution with increasingly realistic physics and to confront the model predictions with future ALMA Zeeman observations of the field strength in disks of different evolutionary stages.
\end{enumerate}

\section*{Acknowledgements}

We thank the referee, Chris McKee, for constructive comments. KHL is supported in part by a graduate fellowship from NRAO through an ALMA SOS (Student Observing Support) award. CYC and ZYL acknowledge support from NSF AST-1815784. ZYL is supported in part by NASA 80NSSC18K1095 and NSF AST-1716259 and 1910106. KT acknowledges support by the Ministry of Education, Culture, Sports,
Science and Technology (MEXT) of Japan, Grants-in-Aid for Scientific Research, 16H05998 and 16K13786.




\input{main.bbl}








\bsp    
\label{lastpage}
\end{document}